\documentclass[onecolumn,authoryear]{els-mrw} 

\usepackage{amsmath,amssymb,amsfonts,amsthm,makeidx,graphicx}
\usepackage{txfonts}
\usepackage{helvet}
\usepackage{graphicx} 
\usepackage{xcolor}
\usepackage{amsmath}
\usepackage{hyperref}
\usepackage{gensymb}
\usepackage[utf8]{inputenc}

\begin{document}

\chapter{Short-range correlations in nuclei}

\author[1]{Or Hen}%
\author[2]{Holly Szumila-Vance}%
\author[3]{Lawrence Weinstein}%

\address[1]{\orgname{Massachusetts Institute of Technology}, \orgdiv{Department of Physics}, \orgaddress{Cambridge, MA 02139, USA}}
\address[2]{\orgname{Florida International University}, \orgdiv{Department of Physics}, \orgaddress{Miami, FL 33199, USA}}
\address[3]{\orgname{Old Dominion University}, \orgdiv{Department of Physics}, \orgaddress{Norfolk, VA 23529, USA}}

\articletag{Chapter Article tagline: update of previous edition,, reprint..}

\maketitle

\begin{glossary}[Glossary]
\term{nucleon} proton or neutron \\
\term{isospin} isomorphic to spin, protons are isospin $+1/2$ and neutrons are isospin $-1/2$
\end{glossary}

\begin{glossary}[Nomenclature]
\begin{tabular}{@{}lp{34pc}@{}}
SRC & Short-range correlations\\
$NN$ & nucleon-nucleon\\
$np$ & neutron-proton \\
$pp$ & proton-proton \\
\end{tabular}
\end{glossary}

\begin{abstract}[Abstract]
Atomic nuclei are held together by the strong nuclear force acting between protons and neutrons (nucleons). While the long-range, averaged part of this force is well described by the nuclear shell model, the short-range and tensor components create a fascinating sub-structure: pairs of nucleons that momentarily approach each other very closely, acquiring large relative momenta. These short-range correlated (SRC) pairs account for roughly 20\% of all nucleons in any nucleus and  almost all of the  high-momentum nucleons. This chapter provides an introduction to SRC pairs: their origin in the nucleon-nucleon tensor force, the experimental methods used to study them — principally deep inelastic and quasi-elastic electron and proton scattering — and the comprehensive picture that has emerged over the past three decades. 
\end{abstract}

\section{Introduction}
The atomic nucleus is on the order of $10^{-15}$~m (1 fm) in diameter and is composed of protons and neutrons bound together by the strong nuclear force. Understanding nuclear structure such as how nucleons move, interact, and organize themselves is one of the central challenges of modern physics.

The simplest model of the nucleus is the Fermi Gas Model in which nucleons fill quantum states up to a maximum momentum (the Fermi momentum, $k_F$).  At nuclear saturation densities, this gives a typical Fermi momentum $k_F\approx 200-250$ MeV/c for medium to heavy nuclei, see~\cite{moniz71}.  While this model establishes the typical nucleon momentum scale, it does not provide information about nuclear structure.

\begin{figure}[htb]
\begin{center}
\includegraphics[width=1.5in]{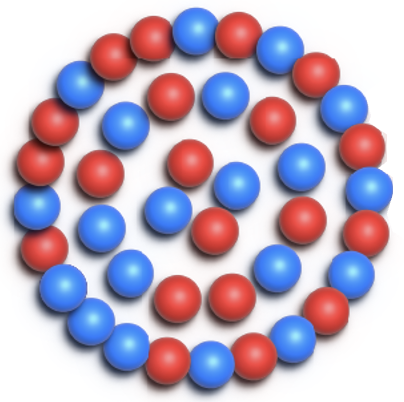}
\caption[]{\label{fig:orbital-cartoon} A cartoon of the nucleon orbitals in $^{40}$Ca showing two protons and two neutrons in the $1s$ orbital, six protons and six neutrons in the $1p$ orbital, and 12 protons and 12 neutrons in the $2s1d$  orbital. Figure courtesy Jefferson Lab.}  \end{center}
\end{figure}

For more than half a century, the independent-particle shell model has been the cornerstone of nuclear theory, see~\cite{Caurier:2004gf}. In this model, each nucleon moves independently in a smooth average (mean-field) potential created by all the other nucleons (see Fig.~\ref{fig:orbital-cartoon}), occupying well-defined quantum orbits organized in shells in direct analogy with electron shells  in atoms. The shell model has been remarkably successful, explaining the special stability of nuclei with ``magic numbers" of protons or neutrons, predicting ground-state spins and parities, and guiding the interpretation of other nuclear properties.

\begin{figure}[htb]
\begin{center}
\includegraphics[width=2.5in]{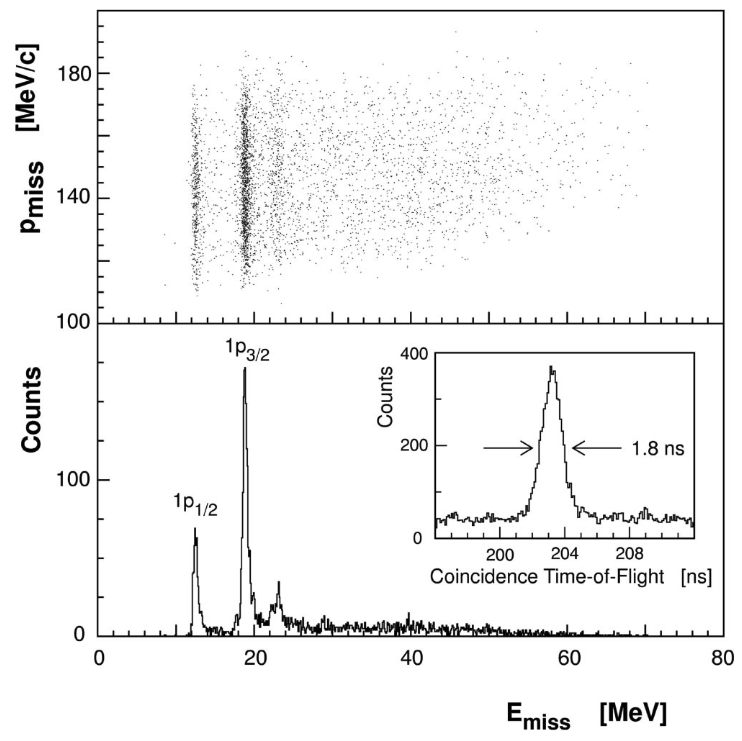}
\includegraphics[width=2.5in]{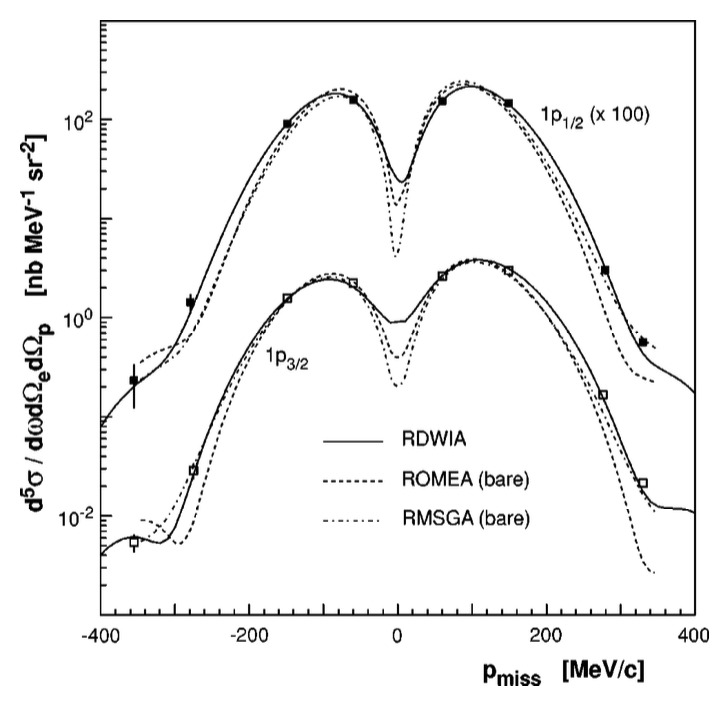}
\caption[]{\label{fig:OeepEmissPmiss} The number of counts in the O$(e,e'p)$ reaction at $Q^2=0.8$ GeV$^2$ (left,top) plotted as a function of missing momentum ($P_{miss}$) versus missing energy ($E_{miss}$); (left,bottom) plotted versus missing energy.  The inset shows the relative time of the electron and proton leaving the target; (right) The O$(e,e'p)$ cross section at $Q^2=0.8$ GeV$^2$ plotted versus missing momentum for protons knocked out of $1p_{1/2}$ and $1p_{3/2}$ shells.  The curves correspond to different calculations including the effects of knocked-out proton reinteraction with the residual nucleus. Figure adapted from~\cite{fissum04}.}
\end{center}
\end{figure}

This shell structure can be seen in $A(e,e'p)$ electron scattering experiments where both the scattered electron and knocked-out proton are detected, see~\cite{kelly96}.  In this case the missing energy $E_{miss}=\omega-T_p-T_{A-1}$ (where the energy transfer $\omega = E-E'$, $E$ and $E'$ are the incident and scattered electron energies and $T_p$ and $T_{A-1}$ are the kinetic energies of the knocked out proton and residual nucleus) and missing momentum $\vec p_{miss}=\vec p - \vec q$ (where $\vec p$ is the outgoing proton momentum and the momentum transfer $\vec q=\vec k - \vec k'$ and $\vec k$ and $\vec k'$ are the incident and scattered electron momenta) are related to the proton binding energy and initial momentum.  Fig.~\ref{fig:OeepEmissPmiss}(left) shows the O$(e,e'p)$ missing energy distribution, showing distinct peaks for proton knockout from $1p_{1/2}$ and $1p_{3/2}$ shells and a much broader distribution at $30\le E_{miss}\le 50$ MeV corresponding to s-shell knockout.  Fig.~\ref{fig:OeepEmissPmiss}(right) shows the O$(e,e'p)$ missing momentum distributions, showing the characteristic $l=1$ momentum distributions of $1p$ orbitals, see~\cite{fissum04}.

\begin{figure}[htb]
\begin{center}
\includegraphics[width=2in]{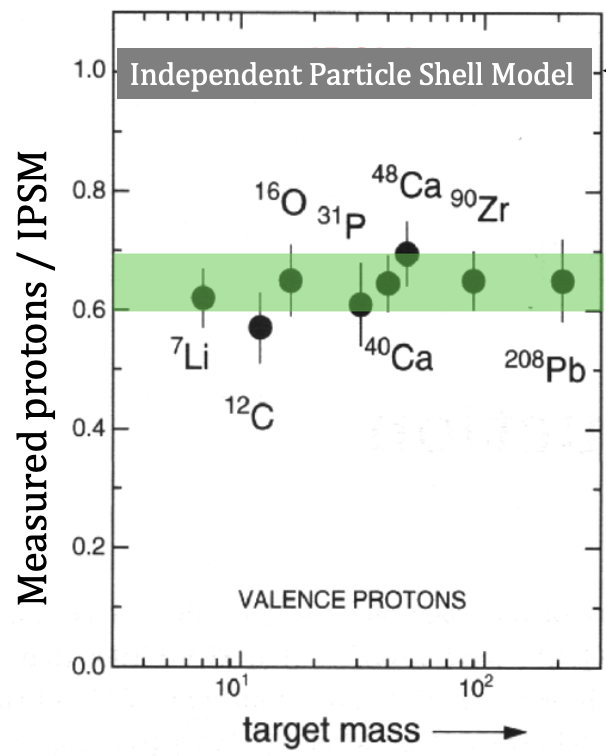}
\caption[]{\label{fig:lapikas-protons} Valence proton spectroscopic strengths were extracted for several nuclei from the ratios of  $A(e,e'p)$ cross section measurements to mean-field  independent particle shell model  calculations. Figure adapted from ~\cite{lapikas97}.}
\end{center}
\end{figure}

However, while the nuclear shell structure was clearly seen,  only 60–70\% of the expected number of protons were found in each shell, see~\cite{kelly96,lapikas97} and Fig.~\ref{fig:lapikas-protons}. 
Even with modern long-range correlation effects included, this only accounts for 80\%, see~\cite{Dickhoff:2004xx} of the measured protons. Something is missing from our shell model description. That missing something is the subject of this chapter: short-range correlated (SRC) nucleon pairs, created by the short-range and tensor components of the nuclear force that lie beyond the mean-field (shell model) description.

The experimental study of SRC pairs has matured significantly since the early theoretical predictions of Frankfurt and Strikman in the 1980s, see~\cite{Frankfurt:1988nt}. Today, precision measurements at major accelerator facilities have established a detailed and quantitative picture of SRC pairs for a range of nuclei. In parallel, deep inelastic scattering (DIS) experiments reveal that the quark-gluon structure of nucleons themselves are modified when inside the nucleus (the ``EMC effect"), and there are compelling indications that this modification is connected to SRC pairs.

\section{The Nuclear Force and the Origin of Short-Range Correlations}
\subsection{The Nucleon-Nucleon Potential}
\begin{figure}[htb]
\begin{center}
\includegraphics[width=2.5in]{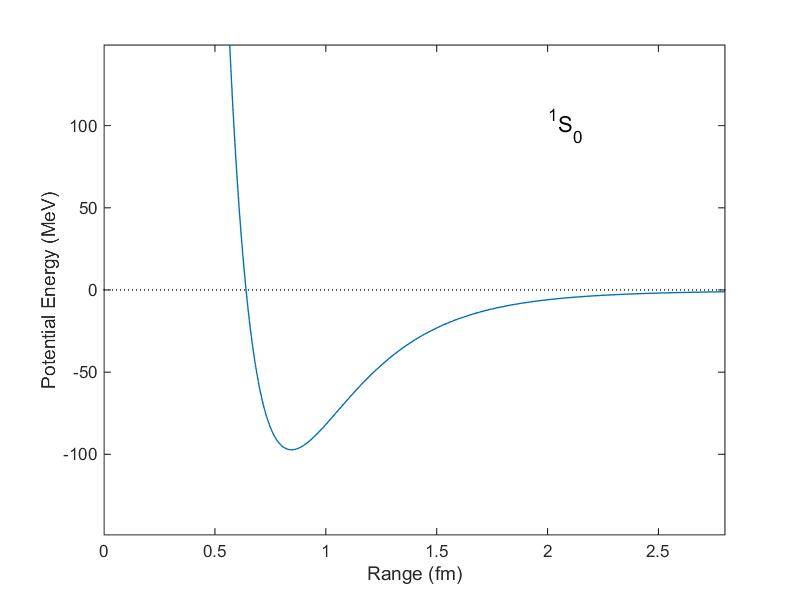}
\caption[]{\label{fig:NNpotential} The central part of the nucleon-nucleon potential $V(r)$ as a function of inter-nucleon separation range $r$. At short range ($r < 1$~fm) the interaction is strongly repulsive, transitioning to an attractive potential at intermediate and long ranges (1–3~fm). Figure from~\cite{ReidPotentialWiki}, licensed under CC BY-SA 4.0.}
\end{center}
\end{figure}

To understand nucleon interactions at short-range, we must first explore how the nucleon-nucleon  interaction varies with the inter-nucleon separation, $r$ (see Fig.~\ref{fig:NNpotential}). 
The \textit{long range} component of the nucleon-nucleon (NN) potential ( $r>2$~fm) is weakly attractive and dominated by the one-pion exchange (OPE) description. The \textit{intermediate range} ($r\approx 0.8-1.5$~fm) is strongly attractive with a potential energy minimum at $r\approx0.8$~fm.  The \textit{short range} component ($r<0.8$~fm) is strongly repulsive. Nucleons resist being compressed too closely together, thereby preventing collapse of the nucleus. 

In addition to the  central (scalar) interaction, there is a smaller tensor component that depends on $S$, the total spin of the nucleon pair.  However, near  the potential minimum where the central force is small,  the tensor force is more important. This  region lies between the long-range attraction and short-range repulsive core, at distances consistent with short-range correlated pair distributions. 

\subsection{What are Short-Range Correlated Pairs?}
Short-range correlated pairs arise when two nucleons momentarily approach each other closely, so that their mutual interaction far exceeds their  interactions with the other $A-2$ nucleons (\cite{Frankfurt88,ciofi15}). For this brief instant, the pair is effectively isolated, experiencing only each other’s force. %
\begin{figure}[htb]
\begin{center}
\includegraphics[width=2in]{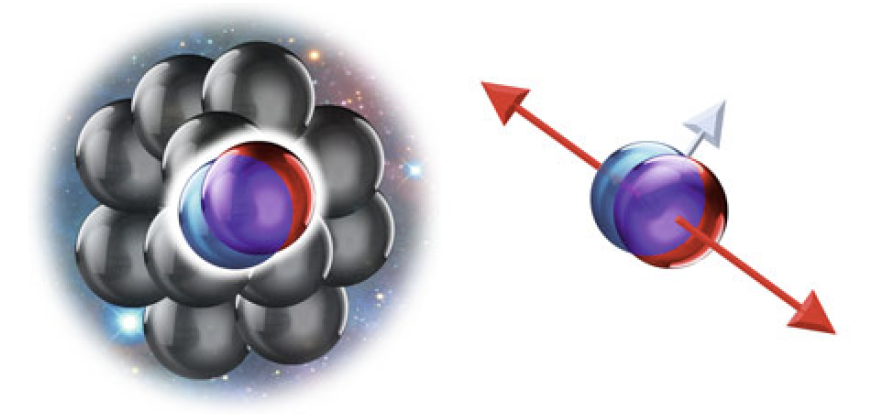}
\caption[]{\label{fig:src-cartoon} Schematic of a short-range correlated pair in coordinate space (left) and momentum space (right). Red arrows show the relative momentum $\vec{p_{rel}}=\vec{p_1}-\vec{p_2}$, and the white arrow shows the center of mass momentum $\vec{p_{cm}}=\vec{p_1}+\vec{p_2}$. Figure from~\cite{Piasetzky2020}.}
\end{center}
\end{figure}
In coordinate space, SRC pairs are nucleon pairs with separation $r\le 1$~fm, much smaller than the average inter-nucleon spacing of about 1.7 fm. In momentum space, they are pairs whose relative momentum $\vec p_{rel}=\dfrac{1}{2}(\vec p_1 - \vec p_2)$ exceeds  $k_F$, while their center-of-mass (CM) momentum $\vec p_{CM}=\vec p_1+\vec p_2$ is the same as it was before their hard interaction, see Fig.~\ref{fig:src-cartoon}.  In order to have large $p_{rel}$ and smaller $p_{CM}$, the individual nucleon momenta, $\vec p_1$ and $\vec p_2$, must point in nearly opposite directions.   This back-to-back momentum-space configuration  is the experimental signature used to identify SRC pairs.

\subsection{SRCs and High momentum}

\begin{figure}[htb]
\begin{center}
\includegraphics[width=3in]{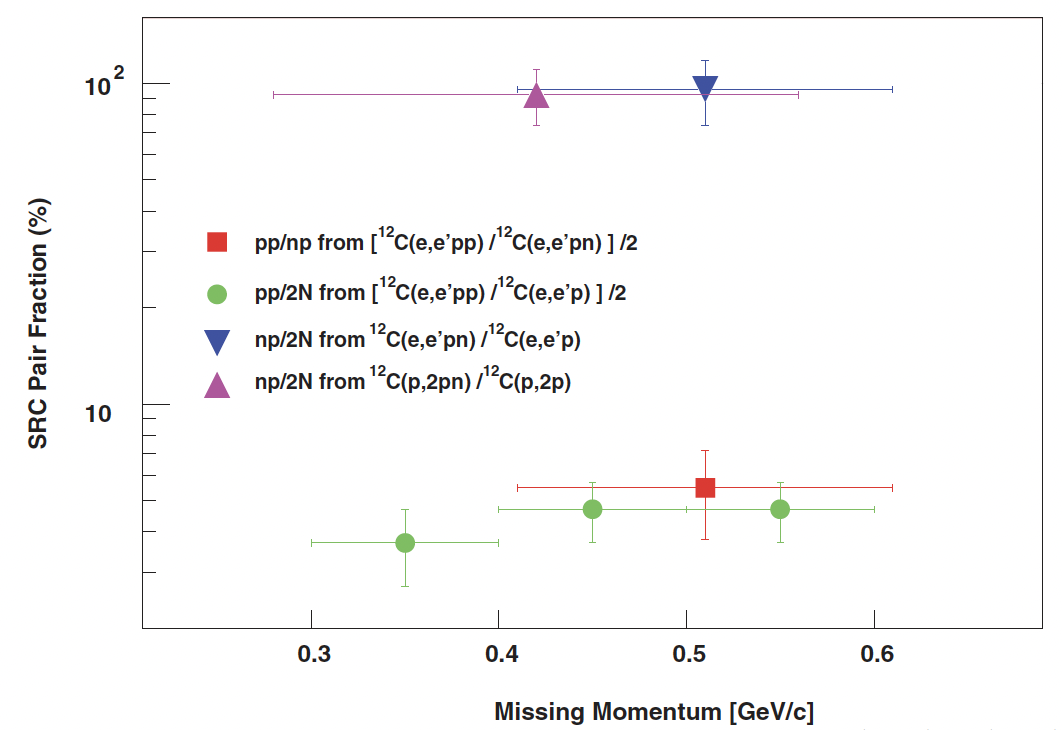}
\caption[]{\label{fig:np-subedi} Short-range correlated pair fractions in carbon as a function of missing momentum. Red squares represent the $pp$-to-$pn$ ratio extracted from the $(e,e'pp)$ to $(e,e'pn)$ cross section ratio. Plotted against $p_{miss}$ are the $np$ pair fractions extracted from $(e, e'pn)/(e, e'p$) (blue triangles)
and from ($p, 2pn$)/($p, 2p$) (magenta triangles), together with the $pp$ pair fraction extracted from ($e, e'pp)/(e, e'p$) (green circles). Figure from \cite{subedi08} with data also from~\cite{tang03}.}
\end{center}
\end{figure}

Two main features of SRC pairs are that almost all high-momentum nucleons in the nucleus belong to an SRC pair and almost all of these pairs are $np$ pairs.   Measurements of proton knockout using both proton (\cite{tang03}) and electron (\cite{subedi08}) probes showed that knockout of a high-$p_{miss}$ proton was almost always associated with the backward emission of a 2nd nucleon and that 2nd nucleon was almost always a neutron, see Fig.~\ref{fig:np-subedi}.  The cause and consequences of $np$ dominance will be explored later.  

\subsection{Observables}
SRC studies seek to understand the properties of SRCs in terms of pairing probabilities, momentum distributions and spectral functions (joint energy momentum distributions).  However, most of these are not observables, but are related to measured quantities by non-unique convolutions (\cite{Furnstahl:2010wd}). This problem is very similar to that of parton distribution functions (PDFs) in high energy physics. Typically the cross section is expressed in terms of a short-distance, calculable piece and a long-distance part expressed as PDFs. This works because a) a rigorous theoretical frame work was developed for PDFs and related functions and b) the same PDFs  appear in all reactions such as deep inelastic lepton scattering, Drell-Yan, and jet production.

For example, both hard and soft nucleon-nucleon potentials can be used in calculations.  The softening of phenomenological and effective field theory (EFT) potentials by renormalization group (RG) transformations that decouple low and high momenta makes calculations of nuclear ground states far easier.  Calculations of high-momentum quantities require a corresponding transformation of the operators, leading to a decreasing hierarchy of many-body forces, see~\cite{Bogner:2009bt}. 

Therefore two ingredients are required to put nuclear SRC pair studies on the same sound footing as quark PDF studies, experimental measurements with a variety of probes (electrons, proton, and photons) and theoretical agreement on the decomposition of calculations. 

\section{Experimental Methods for Studying Short-Range Correlations}
\subsection{Hard Scattering and Electrons}
To study SRC pairs experimentally, physicists scatter high-energy probes, typically electrons or protons, off atomic nuclei and detect the particles that emerge. One challenge is distinguishing the signature of SRC pairs from the many other processes that can produce similar-looking events, see~\cite{frullani84,kelly96,Hansen:2003sn,Cosyn:2009bi,ciofi15}.

``Hard" scattering reactions are characterized by high energy and momentum transfer. This provides scale separation between the energy and momentum of the interaction and the energy and momentum of the nucleon in the nucleus.  When a high-energy electron scatters off a single nucleon in the nucleus, it can transfer enough momentum to knock the nucleon out of the nucleus. If this nucleon belongs to an SRC pair, its partner nucleon will also be ejected. By detecting one or both of the nucleons in coincidence with the scattered electron, one can directly observe SRC pairs.

\begin{figure}[htb]
\begin{center}
\includegraphics[width=2in]{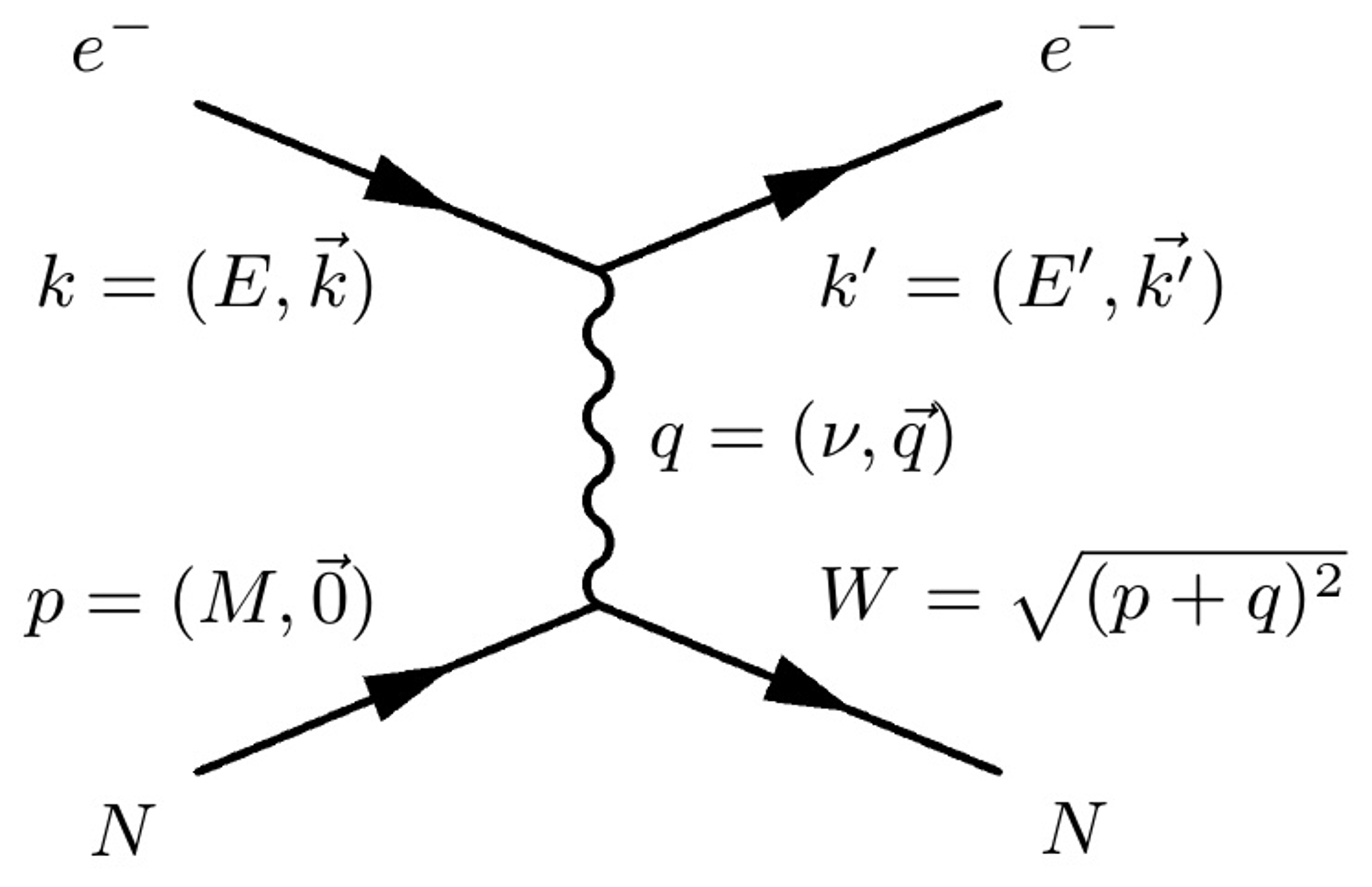}
\caption[]{\label{fig:ope-feynman} Reaction diagram depicting an electron scattering off a nucleon via exchange of a single virtual photon $\gamma^*$ carrying energy and momentum $q = k - k' = (\nu, \vec{q}\thinspace)$. $W$ is the invariant mass of the virtual photon plus nucleon at rest.  Here $M$ is the nucleon mass, elsewhere in this manuscript as $m$.}
\end{center}
\end{figure}

The advantage of using electrons as the probe is that the electron interacts with the nucleus through the well-understood and weak electromagnetic force.   
This means that a) the interaction samples the entire nucleus and not just the surface, b) the interaction is dominated by single photon exchange (1st Born approximation), and c) the electron interacts only once in the nucleus.  

We typically characterize inclusive electron scattering  using 
\begin{itemize}
    \item $Q^2=\vec{q}\thinspace^2-\nu^2$: The four-momentum transfer squared of the reaction where  $q$ is the three-momentum transfer and $\nu$ is the energy transfer. 
    \item $x_B=Q^2/(2m\nu)$: The Bjorken scaling variable where $m$ is the nucleon mass. For deep inelastic scattering, $x_B$ represents the momentum fraction of the struck quark.  For elastic scattering from a free proton at rest, $x_B=1$. For quasielastic  scattering from nucleons in a nucleus, $x_B$ and $Q^2$ determine the minimum initial momentum of the struck nucleon.
\end{itemize}

\subsection{Selecting SRCs in Kinematics \label{sec:srckin}}
In the Impulse Approximation for quasi-elastic (QE) scattering, the virtual photon with four-momentum $q_{\mu}=(\vec{q},\nu)$ is absorbed by a single nucleon, which is knocked out of the nucleus without excitation  (see Fig.~\ref{fig:ope-feynman}).  In the simplest picture (the Plane Wave Impulse Approximation, PWIA), the knocked-out nucleon does not rescatter as it leaves the nucleus.  In this case, the proton-knockout cross section factorizes into an elementary cross section times the spectral function:\\
\begin{equation}
    \dfrac{d\sigma}{d\nu d\Omega_e dE_{miss} d\Omega_p} = K\sigma_{ep}S(E_{miss},p_{miss})
\end{equation}
in which the spectral function $S(E_{miss}, p_{miss})$ encodes the joint probability of removing a bound proton with initial momentum $p_{miss}$ and separation energy $E_{miss}$. The remaining ingredients are the off-shell $ep$ cross section $\sigma_{ep}$ (see, e.g., the de Forest CC1/CC2
prescription from~\cite{deforest83}), an overall kinematic factor $K$, and the phase space of the scattered lepton ($\Omega_e$) and ejected proton ($\Omega_p$). The missing momentum  and missing energy are:
\begin{eqnarray}
&E_{miss} = \nu-T_p-T_{A-1}, \cr
&\vec{p}_{miss} = \vec p_p - \vec q 
\end{eqnarray}
where $T_p$ and $T_{A-1}$ are the kinetic energies of the detected proton and residual nucleus, respectively.   Some papers define $\vec{p}_{miss}= \vec q - \vec p_p$.  If the scattered proton rescatters as it exits the nucleus (final state interactions or FSI), then relationship between $p_{miss}$ and the proton initial momentum is muddied and the cross section no longer exactly factorizes. While the spectral function itself is not an observable, it is extracted from the measured cross section in a high-momentum, one-body operator picture.  See~\cite{Furnstahl:2010wd,Bogner:2009bt} for more discussion.

\begin{figure}[htb]
\begin{center}
\includegraphics[width=2in]{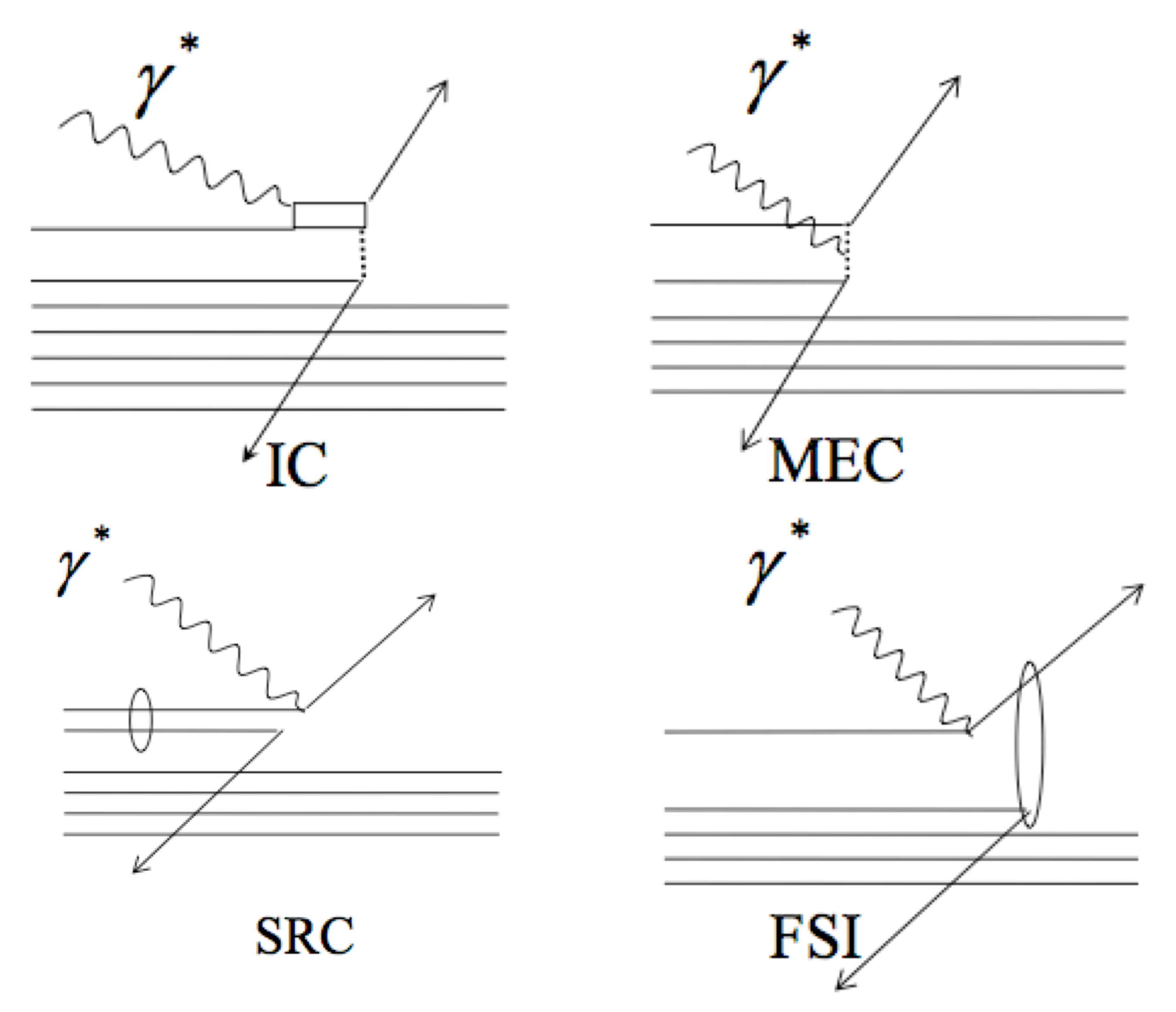}
\caption[]{\label{fig:srcbg-feynman} Two nucleon emission processes.  SRC refers to the absorption of the virtual photon on one nucleon of an SRC pair; IC refers to virtual photon absorption on a single nucleon, exciting it to a resonance, which then de-excites; MEC refers to virtual photon absorption on a meson being exchanged between two nucleons; FSI refers to rescattering of the knocked out-proton by another nucleon.  All of these result in the emission of two nucleons.}
\end{center}
\end{figure}

We typically use QE $(e,e'p)$ to study single-nucleon energy and momentum distributions in the nucleus.  However, when we study nucleons in short range correlated pairs, we are considering reactions where two nucleons are emitted.  There are several processes that can result in two-nucleon emission, see  Fig.~\ref{fig:srcbg-feynman}:
\begin{itemize}   
\item\textbf{Short Range Correlated Pairs (SRCs)} The virtual photon is absorbed on one nucleon of the correlated pair, knocking it out.  The high-momentum spectator correlated nucleon also is emitted from the nucleus.

\item\textbf{Meson Exchange Currents (MEC)}: The virtual photon can be absorbed by a pion being exchanged between two nucleons. This often leads to two-nucleon emission. MEC effects are suppressed at large $Q^2$ because the electromagnetic form factor of the pions decreases faster with momentum transfer than the nucleon form factor.

\item\textbf{Isobar Configurations (IC)}: The photon can excite a nucleon to a $\Delta$ or $N^*$ resonance, a heavier excited state of the nucleon, which then reinteracts with a second nucleon, leading to emission of both nucleons. This requires greater energy transfer than quasi-elastic scattering. The cross section for IC decreases faster with $Q^2$ than QE scattering.  Requiring $x_B>1$ (smaller energy transfer) also suppresses this contribution.

\begin{figure}[htb]
\begin{center}
\includegraphics[width=2.6in]{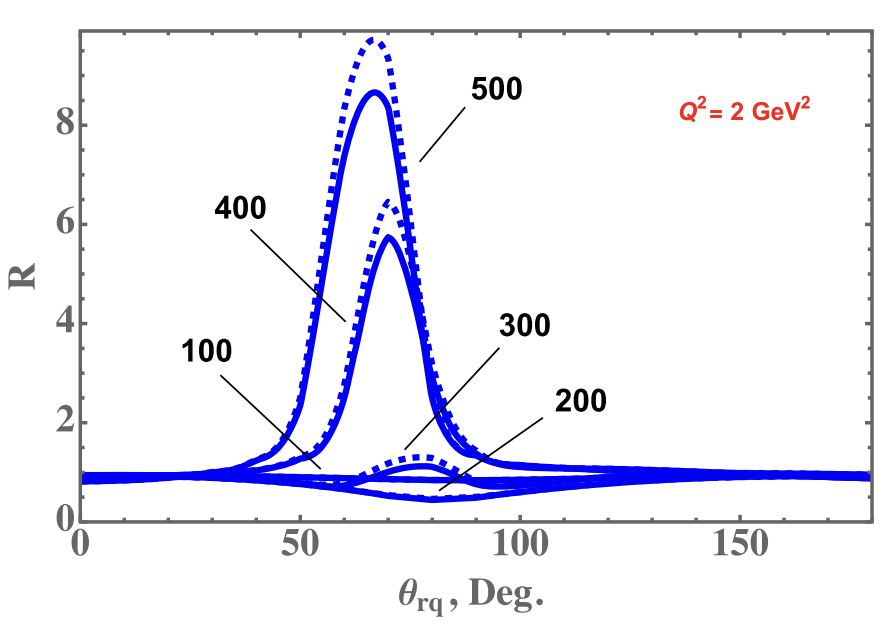}
\caption{\label{fig:rescatter-sargsian} {The ratio of the full $d(e,e'p)$ cross section calculation including FSIs divided by the cross section in the plane wave impulse approximation (no FSIs),  $R=\dfrac{\sigma_{PWIA+FSI}}{\sigma_{PWIA}}$, for different missing momenta, plotted versus $\theta_{rq}$,  the angle between the recoiling spectator nucleon and $\vec{q}$. The different curves are shown for  different  $p_{miss}$ (in MeV/c) where the solid lines correspond to the full calculations, and the dashed lines also include the effects of single charge exchange interactions. Figure from~\cite{BOEGLIN2024}.}}
\end{center}
\end{figure}

\item\textbf{Final State Interactions (FSI)}: After being struck by the photon, the outgoing nucleon can rescatter off another nucleon on its way out of the nucleus. This  changes the momentum of the detected nucleon. There are two general possible results: 
\begin{itemize}
\item The detected nucleon can rescatter slightly, typically increasing the missing momentum.  This lowers the cross section for low-$p_{miss}$ nucleons ($p_{miss}\le 250$ MeV/c) and increases the cross section for high-$p_{miss}$ nucleons.  This small angle rescattering gives a peak at angles between the recoil momentum and $\vec q$ around $\theta_{rq}\approx 70^\circ$.  These events are typically cut by requiring smaller $\theta_{rq}$.  See Fig.~\ref{fig:rescatter-sargsian}.
\item The detected nucleon can have a much harder rescattering (including inelastic rescattering) and is not detected.  This proton ``absorption" is typically calculated using the Glauber approximation at outgoing proton momenta $p_p\ge 500$ MeV/c.  These calculations are accurate to about 10\%.
\end{itemize}
\end{itemize}

\begin{figure}[htb]
\begin{center}
\includegraphics[width=2.5in]{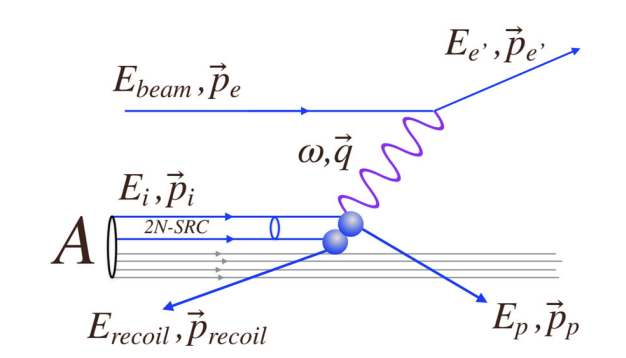}
\caption[]{\label{fig:src-feynman} The virtual photon interacts with a nucleon from a SRC pair, and both interacting nucleons are ejected.  The energy transfer here is indicated by $\omega$ rather than $\nu$. Reprinted figure with permission from \cite{Cohen:2018gzh}. Copyright 2018 by the American Physical Society.}
\end{center}
\end{figure}

To restrict the scattering reaction to  SRC pairs as shown in Fig.~\ref{fig:src-feynman}, the general kinematic requirements are:
\begin{itemize}
    \item $Q^2>1.5$~GeV$^2$: ensures hard scattering and suppresses the meson-exchange current (MEC) contributions
    \item $x_B>1.2$: selects nucleons with initial momentum greater than the Fermi momentum and suppresses the excitation of nucleon resonances called isobar configurations (IC)
    \item $\theta_{rq}<40\degree$: the angle between the recoiling nucleus and the momentum transfer $\vec{q}$ should be small such that the recoiling system moves forward. This minimizes in-scattering from final state interactions (FSI), which are maximum at $\theta_{rq}\sim70\degree$ (\cite{BOEGLIN2024}). See Fig.~\ref{fig:rescatter-sargsian}. 
    \item $|\vec{p}_{miss}|>0.3$~GeV/$c$: selects  nucleons with momentum above the Fermi momentum, identifying high-momentum nucleons that are likely to be in SRC pairs
\end{itemize}

\subsection{Other probes}
In order to show that SRCs are a fundamental part of nuclear structure and to put SRC studies on the same rigorous level as parton PDFs, SRC studies need to be probe independent.

\begin{figure}[htb]
\begin{center}
\includegraphics[width=2in]{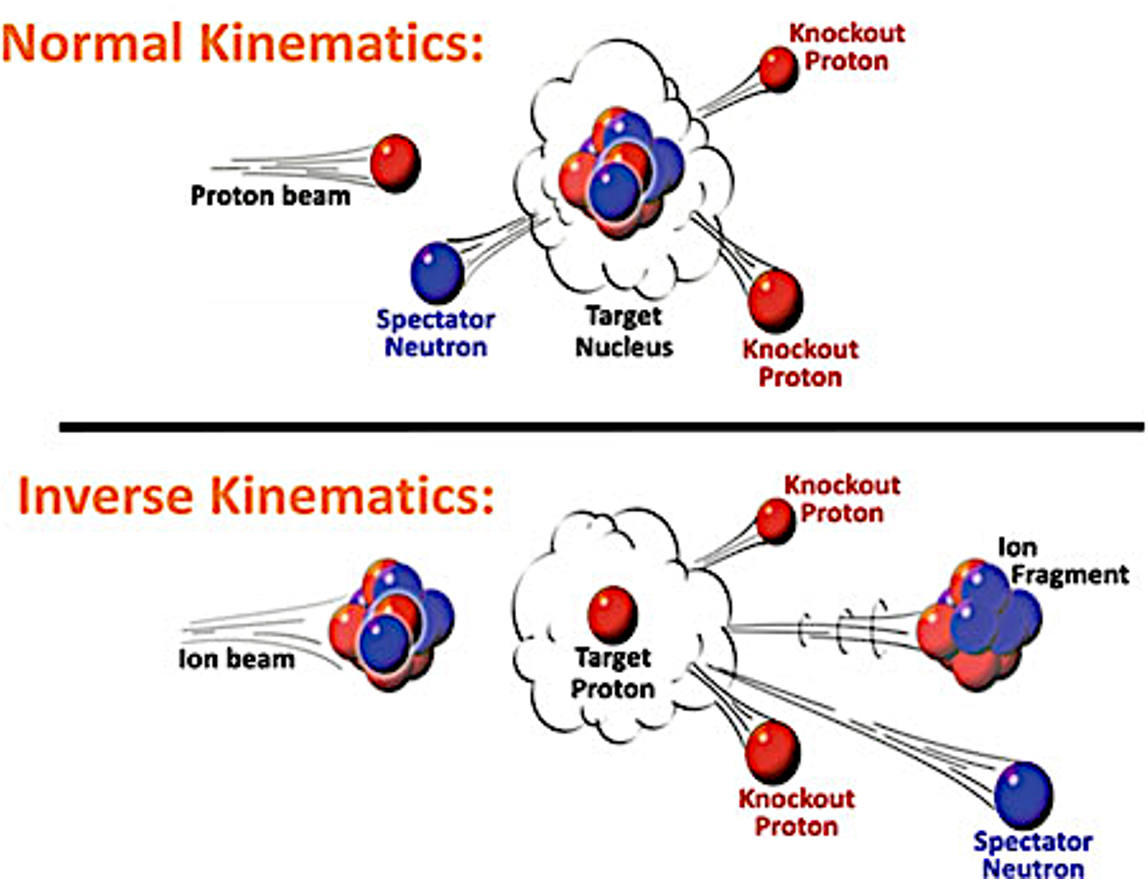}
\caption[]{\label{fig:invkin-cartoon} Diagrams illustrating the study of SRC pairs in nuclei via $pA$ reactions in normal (top) and inverse (bottom) kinematics. Protons are represented in blue and neutrons in red. In inverse kinematics, the residual nucleus (ion fragment) can be detected alongside the knocked-out and spectator nucleons, providing additional experimental constraints. Figure from \cite{Patsyuk:2021jea}.}
\end{center}
\end{figure}

This was first shown using  proton beams. In proton-nucleus scattering, one proton in the beam scatters quasi-elastically off a proton in the target nucleus ($pp\rightarrow pp$ hard scattering). If the struck proton belongs to an SRC pair, its partner nucleon is emitted as well as shown in Fig.~\ref{fig:invkin-cartoon}.  Because high-energy proton-proton scattering decreases as $s^{-10}$ at center-of-mass scattering angles of $\approx 90^\circ$, $pp$ quasielastic scattering will preferentially select nuclear protons with momentum in the beam direction.  

A proton-carbon scattering experiment at Brookhaven National Lab measured two outgoing protons and a possible backward neutron.  They found that high-momentum neutrons ($p_n\ge k_F=220$ MeV/c) were emitted backward from the missing momentum, while low momentum neutrons were emitted isotropically, indicating that high-momentum backward neutrons were the correlated spectators in $np$ SRC pairs, see~\cite{tang03,piasetzky06,Aclander:1999fd}.  They further found that $92^{+8}_{-18}\%$ of high-$p_{miss}$ protons had a correlated neutron partner (\cite{piasetzky06}), consistent with later electron-scattering results (see Fig.~\ref{fig:np-subedi}).

Proton scattering experiments can also be performed in inverse kinematics, where a beam of heavy ions (such as $^{12}$C) hits a hydrogen target. This allows experimenters to detect and identify the residual nucleus traveling forward in the lab frame. By detecting  a specific  nucleus (e.g., $^{10}$B or $^{10}$Be in the  $p(^{12}\mathrm{C},ppA)$ reaction), one can identify the type of SRC pair that was broken up, significantly reducing background from other processes. This process is illustrated in the bottom of Fig.~\ref{fig:invkin-cartoon}.  It also allows studies of nuclei too unstable to used in a conventional fixed target experiment.

Similarly, one can also perform experiments with real photons.  In this case, energy and momentum conservation forbids quasielastic scattering.  However, one can instead study quasielastic meson photoproduction reactions.

\section{Key Experimental Measurements and Results}
\subsection{First Evidence: Almost All High-Momentum Protons Have Neutron Partners}

The first measurement of SRC-pair knockout  scattered 6--9 GeV/c protons from C at Brookhaven National Lab, measuring the two outgoing protons at  90$\degree$ in the center of mass frame, see~\cite{tang03,piasetzky06}. As mentioned above,  low-momentum neutrons ($p_n \le 220$ MeV/c) were emitted isotropically, consistent with being uncorrelated spectators.  High-momentum neutrons ($275\le p_n\le 550$ MeV/c) were all emitted backwards with respect $\vec p_{miss}$, a clear signature of knockout from SRC pairs.  They further found that $92^{+8}_{-18}\%$ of high-$p_{miss}$ protons had a correlated neutron partner, see~\cite{piasetzky06}.  The results are shown in Fig.~\ref{fig:np-subedi} in purple.

The second measurement was from\cite{subedi08} and measured C$(e,e'pN)$ at $Q^2= 2$~GeV$^2$ and $x_B=1.2$ with 4.6~GeV electrons in Hall A of Jefferson Lab. They detected the scattered electron and the knocked-out proton in the High-Resolution Spectrometers and detected the possible correlated spectator nucleon in  a large proton spectrometer followed by a neutron detector. Of the $p_{miss} \geq 300$~MeV/$c$ protons knocked out in that experiment, $96\pm22$\% were tagged in coincidence with a recoiling neutron and only $9.5\pm 2$\% with a recoiling proton.  Thus  the partner of a high-momentum nucleon is overwhelmingly of opposite isospin (e.g., $np$ or $pn$).
Once the single charge exchange contamination is accounted for  (a final state interaction in which an outgoing correlated neutron undergoes a $(n, p)$ reaction in the residual system and is detected as a proton), the $np$ to $pp$ event ratio reaches $9.0\pm2.5$. Accounting for $np$ and $pp$ pair counting, this translates to an $np$ to $pp$ pair ratio of $18\pm5$ as shown in Fig.~\ref{fig:np-subedi}. In other words, $np$ SRC pairs are about 18 times more common than $pp$ pairs.

This remarkable agreement between measurements using different probes (protons and electrons, which interact via different forces) significantly strengthens the conclusion: these are properties of preexisting, high-momentum SRC pairs in the ground state of the nucleus, not artifacts of the reaction mechanism.

\subsection{$np$ Dominance in Heavier Nuclei}
Carbon is a relatively light, symmetric nucleus ($Z=N=6$). Does $np$ dominance persist in heavier and more asymmetric nuclei? The answer is yes.

\begin{figure}[htb]
\begin{center}
\includegraphics[width=2in]{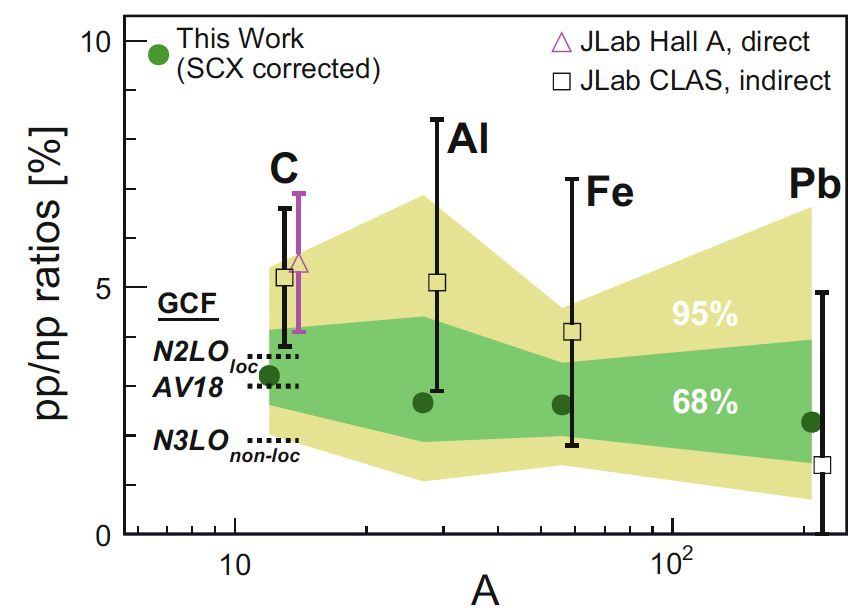}
\caption[]{\label{fig:np-dom-nuclei} Extracted $pp$ to $np$ SRC pair ratios as a function of atomic number $A$. Filled green circles describe the SCX-corrected $pp$ to $np$ cross section ratios, with the shaded
bands giving the 68\% and 95\% credible intervals propagated from the measured cross
sections and the SCX correction. The magenta triangle shows the SCX-corrected carbon result of \cite{subedi08}, and the open black squares are the indirect determination of \cite{hen14}, obtained
from inclusive $a2$ scaling rather than direct two nucleon coincidence. The error bars on both represent the one $\sigma$ confidence limits. Horizontal dashed lines indicate Generalized Contact Formalism-calculated contact ratios for carbon using various $NN$ potentials with contact values fitted to the measured cross section ratios, see Sect.~\ref{sect:GCF}. Reprinted figure with permission from \cite{Duer:2018sxh}. Copyright 2019 by the American Physical Society.}
\end{center}
\end{figure}

Measurements at Jefferson Lab scattered  a 5-GeV electron beam from C, Al, Fe and Pb, detecting the outgoing charged particles in   CLAS (the CEBAF Large Acceptance Spectrometer) to measure $(e,e'p)$, $(e,e'n)$,  $(e,e'pp)$ and  $(e,e'pn)$ events simultaneously, see~\cite{hen14,duer18,Duer:2018sxh}. They looked at both the ratio of $(e,e'pp)/(e,e'p)$ and the ratio of $(e,e'pp)/(e,e'pn)$ and found that only a small fraction of high-$p_{miss}$ protons had a correlated proton partner, with the remainder having neutron partners, confirming $np$ dominance for nuclei spanning the periodic table, see Fig.~\ref{fig:np-dom-nuclei}.

\begin{figure}[htb]
\begin{center}
\includegraphics[width=3.9in]{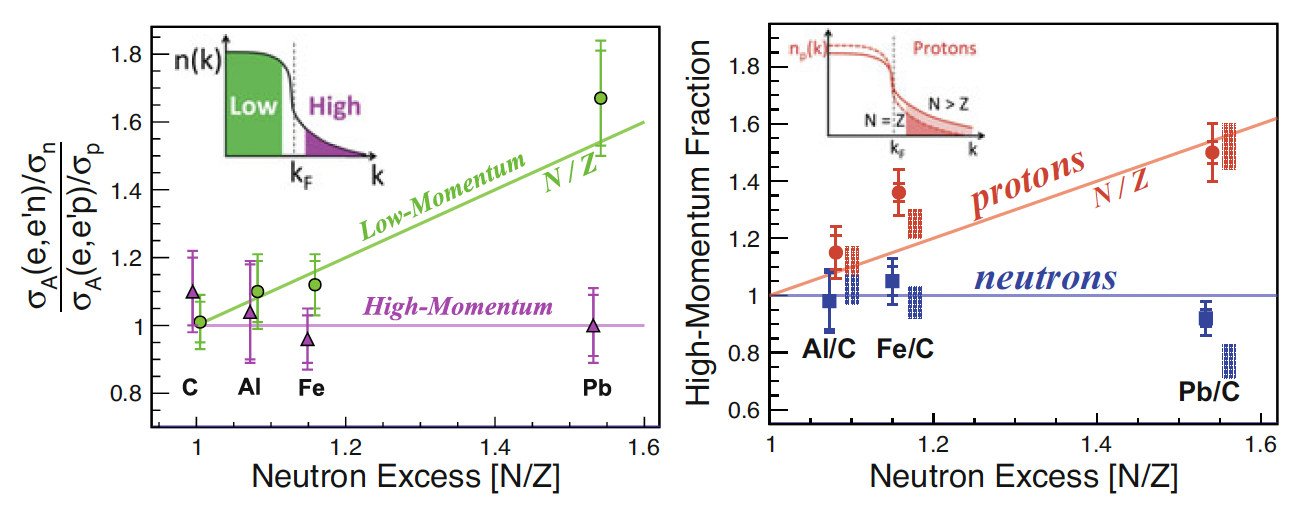}
\caption[]{\label{fig:np-duer} Reduced cross section ratios of $(e,e'n)$ to $(e,e'p)$ events as a function of nuclear mass for nuclei ranging from C to Pb, shown for low missing momentum (green circles) and high missing momentum (magenta triangles), shown on the left. The corresponding curves indicate the expected $N/Z$ scaling at low missing momentum and the predicted ratio of unity at high missing momentum. The inset illustrates the missing momentum ranges considered. On the right is the double ratio of high- to low-missing momentum event yields for nucleus $A$ relative to C, shown separately for protons (red) and neutrons (blue). The horizontal lines at $N/Z$ and one serve as visual guides, while the red and blue shaded bands indicate the range of predictions from the phenomenological $np$ dominance model for protons and neutrons, respectively. Figure from~\cite{duer18}.}
\end{center}
\end{figure}

They further explored the nature of $np$ dominance by extracting the  ratio of $(e,e'n)$ to $(e,e'p)$ reduced cross sections  for both low-$p_{miss}$ (mean field) and high-$p_{miss}$ (SRC) events, see Fig.~\ref{fig:np-duer}a.  The $n/p$ ratio at low-$p_{miss}$ increased as $N/Z$, consistent with simple nucleon counting, while the same ratio at high-$p_{miss}$ was constant at one, consistent with $np$ dominance (if all SRC pairs are $np$ pairs, then the $n/p$ ratio will be $A$-independent).  Both ratios for C are consistent with unity, as expected for a symmetric nucleus.

Similarly, they looked at the high-momentum fractions of neutrons and protons, by forming the high-$p_{miss}$ to low-$p_{miss}$ double ratios of nucleus $A$ to C.  The high-momentum fraction of neutrons is consistent with unity, and the high-momentum fraction of protons increases with $N/Z$, consistent with the additional neutrons forming more SRC $np$ pairs, see Fig.~\ref{fig:np-duer}b.

An important consequence of $np$ dominance in asymmetric nuclei is momentum sharing inversion. In a simple Fermi gas model, adding extra neutrons to a nucleus increases the average neutron momentum (neutrons must fill higher and higher orbits). But in an $np$-SRC description, adding extra neutrons increases the average proton (minority nucleon) momentum, because each extra neutron creates more $np$ SRC pairs, and in each pair both the proton and the neutron acquire high momentum. For a neutron-rich nuclei such as lead (with 126 neutrons and 82 protons), there are more high-momentum protons than one would naively expect, and conversely the high-momentum neutron fraction is diluted. Quantitatively, if 20\% of nucleons are in SRC pairs, then lead has about 40 nucleons in SRC pairs, roughly 20 protons and 20 neutrons. The high-momentum proton fraction is then $20/82\approx24$\%, while the high-momentum neutron fraction is only $20/126\approx$16\%.

\subsection{The Causes of $np$ Dominance}

The dramatic difference between the $np$- and $pp$-pairs, known as ``$np$ dominance", is a direct consequence of the tensor part of the nuclear force.  Since nucleons are fermions, the $NN$ wavefunction must be antisymmetric. SRC pairs are predominantly in orbital angular momentum $L=0$ ($s$-wave)  states, which are spatially symmetric.  $pp$ pairs have    symmetric  isospin\footnote{In isospin space, nucleons have total isospin 1/2, with protons have a $z$-projection of $+1/2$ and neutrons have a $z$-projection of $-1/2$.  The total wavefunction on an $NN$ pair is the product of the spatial, spin and isospin parts.} wavefunctions.  Therefore the spin wavefunction must be antisymmetric: $S=0$, giving total angular momentum 0.  The $L=2$ tensor interaction cannot act on zero angular momentum states.  The $L=0$ wave function for $pp$ pairs has a sharp minimum at $p=400$ MeV/c, see~\cite{schiavilla07}.  $np$ pairs can have an asymmetric isospin wave function and therefore a symmetric spin $S=1$, similar to that of the deuteron.  Since $np$ pairs have total angular momentum 1, they can be acted upon by the $L=2$ tensor part of the $NN$ potential, which fills in the $L=0$ minimum.

\begin{figure}[htb]
\begin{center}
\includegraphics[width=4in]{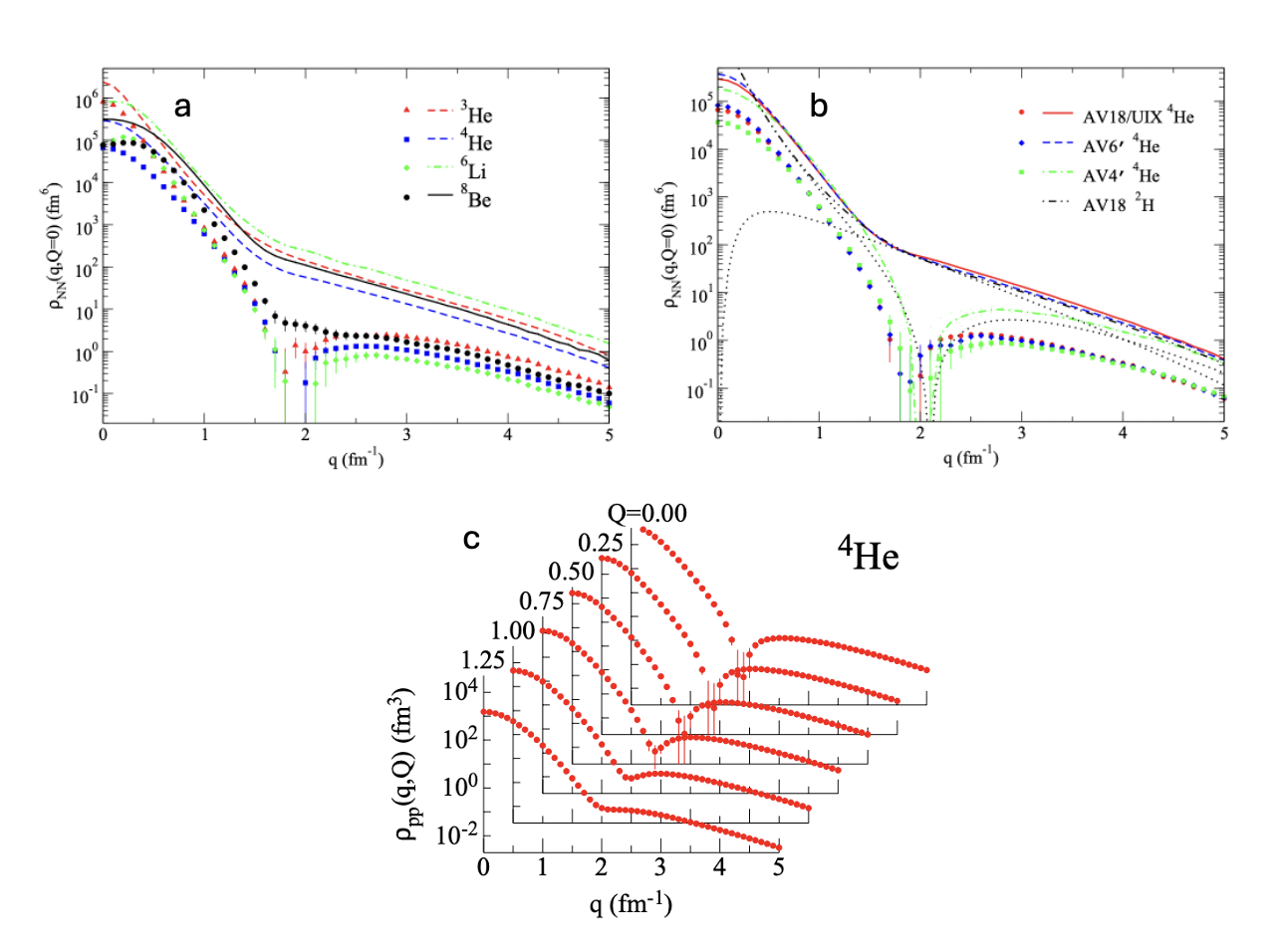}
\caption{\label{fig:anl-NN-calc} {(a) The  relative momentum ($q\equiv p_{rel}$) distributions of $pn$ pairs (lines) and $pp$ pairs (symbols) at center of mass momentum $Q\equiv p_{CM} =0$, for four light nuclei computed with the AV18 nucleon-nucleon interaction: $^3$He (red), $^4$He (blue), $^6$Li (green), and $^8$Be (black). (b) The same $pn$ (lines) and $pp$ (symbols) distributions at $Q = 0$ in $^4$He, compared across different $NN$ potentials of  AV18 (red), AV6' (blue), and AV4' (green) alongside the AV18 result for $^2$H (black). The $s$  and $d$ wave contributions to the deuteron wave function are indicated by the black dashed lines. Both (a) and (b) are reprinted  with permission from \cite{schiavilla07}. Copyright 2007 by the American Physical Society. (c) The $pp$ two-body momentum distribution of $^4$He from variational Monte Carlo (\cite{wiringa14}), integrated over the directions of the relative momentum $\vec{q}$ and the pair total momentum $\vec{Q}$, is shown as a function of $q$ for several slices of $Q$ from 0 to 1.25 fm$^{-1}$. Here, $q$ refers to the relative momentum, $\vec p_{rel}$ in the text, and $Q$ refers to the center of mass momentum $\vec p_{cm}$ in the text. Reprinted figure with permission from \cite{wiringa14}. Copyright 2014 by the American Physical Society.}}
\end{center}
\end{figure}

Fig.~\ref{fig:anl-NN-calc}a shows the relative momentum distributions for $pp$ and $pn$ pairs for different light nuclei using the AV18 $NN$ interaction.  There is a sharp minimum for $pp$ pairs at relative momentum 2 fm$^{-1} \approx 400$ MeV/c, as expected for $l=0$ pairs.  There is no minimum for $np$ pairs. Fig.~\ref{fig:anl-NN-calc}b shows the relative momentum distribution for $pn$ and $pp$ pairs for $^4$He for different $NN$ interactions.  When calculations use a $NN$ potential that deliberately omits the tensor force (AV4'), the $np$ distribution develops the same minimum as the $pp$ distribution, confirming that the tensor force is entirely responsible for $np$ dominance (see \cite{schiavilla07}).

Fig.~\ref{fig:anl-NN-calc}b also shows the $s-$ and $d-$wave contributions to the $^2$H momentum distribution.  The $L=0$ $s-$wave contribution has a sharp minimum at $q=p_{rel}\approx 400$ MeV/c, and the $L=2$ $d$-wave contribution does not have this minimum.  This also shows the tensor nature of the $np$ SRC pairs.

The result is the well known ``$np$ pair dominance" where in any nucleus the tensor channel
populates roughly an order of magnitude more $pn$ SRC pairs than $pp$ or $nn$ SRC pairs. This $np$ dominance has profound implications for the momentum distributions of protons and neutrons in asymmetric nuclei (those with unequal numbers of protons and neutrons).

The minimum at $p_{rel}\approx 400$ MeV/c is deepest when the pair moves slowly  (small $p_{cm}$, labeled $Q$ in these plots), since at vanishing total momentum, the $NN$ pairs are restricted to a relative $s-$wave configuration. At larger center-of-mass momenta, additional partial waves become accessible, and the tensor force contributes to these channels in a way that progressively fills in the minimum, see~\cite{wiringa14}.

\subsection{The Tensor Nature of $np$ Dominance: Evidence from 3He}
Confirmation of the tensor nature of $np$ dominance came from measurements of the $^3$He nucleus (two protons and one neutron). By measuring the fully exclusive $^3$He$(e,e'pp)n$ reaction using the CLAS spectrometer (\cite{niyazov03,baghdasaryan10}), the scattered electron and outgoing protons are detected while the missing neutron is inferred from energy and momentum conservation. 

In these events, the electron knocks one nucleon out, and the remaining $np$ or $pp$ pair is observed. If $np$ dominance is simply about counting (i.e. $np$ pairs outnumber $pp$ pairs because there are more $np$ pairs in $^3$He), we would expect the $pp/pn$ ratio to be roughly $1/2$. Instead, they found the ratio to be much smaller at low pair center of mass momentum, consistent with tensor dominance, and to increase significantly as the pair total center of mass momentum increased.

This behavior is consistent with the theoretical calculations that predicted that at low center of mass momentum, $pp$ pairs must be in $s$-states ($L=0$), where the $pp$ minimum in the momentum distribution is most pronounced. As the CM momentum increases, higher angular momenta become accessible, the tensor interaction can contribute, and the $pp/pn$ ratio increases toward the ratio expected from simple pair counting.

\subsection{Which nucleons pair?}
The question of how SRC pair formation depends on nuclear  characteristics such as  mass, neutron excess, and orbital structure is still open.  Theoretical models will be discussed in a later Section.

\cite{duer18} investigated this by comparing proton 
and neutron knockout from $^{12}$C, $^{27}$Al, $^{56}$Fe, and $^{208}$Pb, nuclei that spanned a wide range of mass and neutron excess (see Fig.~\ref{fig:np-duer}). An increased proton pairing probability was observed in neutron-rich nuclei, suggesting that excess neutrons in asymmetric nuclei form additional $np$ SRC pairs. However, because nuclear mass and neutron excess increased together across the four measured nuclei, the data could not independently constrain their respective contributions to SRC pair formation.

To address this limitation, a subsequent inclusive scattering measurement compared $^{40}$Ca and $^{48}$Ca directly, taking advantage of their identical proton numbers but differing neutron content, see~\cite{Nguyen:2020mgo}. The results suggested that adding 40\% more neutrons to $^{40}$Ca increased the number of SRC pairs by only  $\sim$17\%. However, inclusive scattering has model-interpretation uncertainties making it difficult to draw firm conclusions about the underlying pair-formation mechanism, see~\cite{Weiss:2020bkp}.

Future measurements and forthcoming publications on a wider selection of nuclei will help to resolve the question of how nuclear mass $A$, nuclear asymmetry $N/Z$, and nuclear shell structure each contribute to SRC pairing probability.

\subsection{The transition from Mean Field to SRC}

We know that mean-field nucleons dominate the description of nuclei at low nucleon momentum and that nucleons in SRC pairs dominate at high nucleon momentum.  However, the location and width of the transition region is important.  A narrow transition makes a factorized description of nuclei more accurate.  Inclusive per-nucleon $(e,e')$ cross-section ratios of nucleus $A$ to deuterium (see Section~\ref{sec:a2}) indicate that SRC-based scaling begins at a minimum nucleon momentum of $275\pm25$ MeV/c (\cite{egiyan03}). $A(e,e'p)$ measurements can provide a more detailed picture. 

\begin{figure}[htb]
\begin{center}
\includegraphics[width=3.9in]{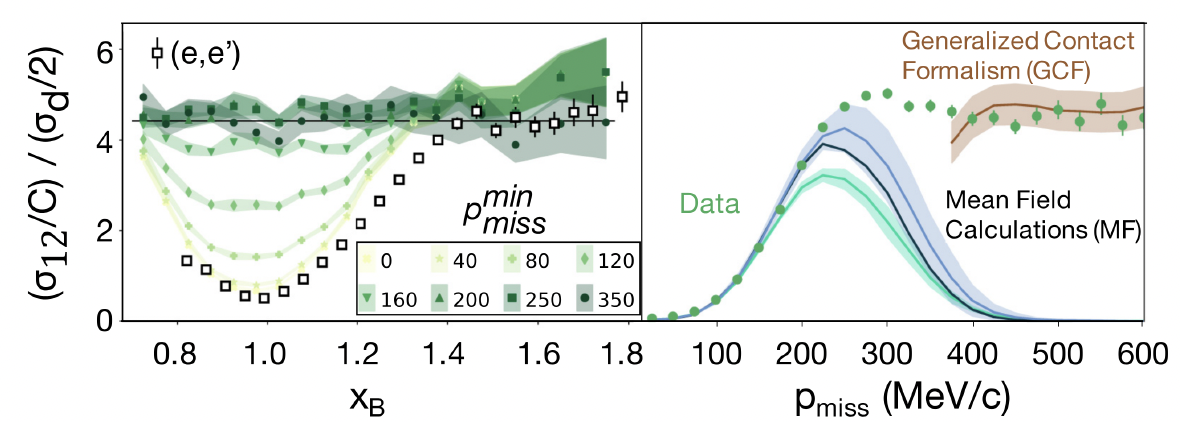}
\caption[]{\label{fig:transition-igor} Per-nucleon $(e,e'p)$ cross-section ratios for carbon to deuterium, shown as a function of $x_B$ (left) and missing momentum (right). (left) Filled symbols represent data integrated over a missing momentum range from $p_{miss}^{min}\leq p_{miss}\leq600$~MeV/$c$, with colored bands indicating total uncertainties at the 68\% confidence level. Open squares show the corresponding inclusive $(e, e')$ per-nucleon cross section ratios for comparison. (right) Cross section ratios integrated over the kinematic range $0.7\leq x_B\leq 1.8$. Green circles denote the measured experimental data. The brown curve shows theoretical cross sections calculated for scattering off nucleons in Short-Range Correlated (SRC) pairs within carbon using the Generalized Contact Formalism (GCF), while the remaining curves correspond to mean field single nucleon calculations from the QMC (teal), IPSM (black), and Skyrme (azure) models. The IPSM and Skyrme results were normalized to the experimental data in the low missing-momentum region below 150~MeV/$c$. Reprinted figure with permission from~\cite{CLAS:2022odn}. Copyright 2023 by the American Physical Society.}
\end{center}
\end{figure}

\cite{CLAS:2022odn} measured per-nucleon $A(e,e'p)$ cross-section ratios for several nuclei relative to deuterium in the Jefferson Lab CLAS spectrometer at $Q^2 > 1.5$~GeV$^2$ as a function of $x_B$, for varying lower bounds on the missing momentum in the range $p_{miss}^{min} \leq p_{miss} \leq 600$~MeV/$c$. At low values of $p_{miss}^{min}$, the semi-exclusive ratios closely resembled their inclusive counterparts, with scaling behavior emerging at $x_B \geq 1.5$. Raising the missing momentum threshold to $p_{miss}^{min} = 350$~MeV/$c$ shifted the onset of scaling to $x_B = 0.7$, demonstrating that the SRC-dominated regime can be isolated by imposing either $x_B \geq 1.5$ or $p_{miss} \geq 350$~MeV/$c$.

To characterize the mean-field to SRC transition, the $x_B$-integrated cross section ratios were examined as a function of $p_{miss}$, see Fig.~\ref{fig:transition-igor}. The ratio increases rapidly up to $p_{miss}\approx 250$ MeV/c and then flattens out, consistent with the onset of SRC scaling seen in inclusive measurements. At $p_{miss}\le 250$ MeV/c, the ratios are well described by mean-field calculations.  At higher $p_{miss}$, the ratios are well described by Generalized Contact Formalism calculations (GCF), see~\cite{CLAS:2022odn} and discussion in Section~\ref{sect:GCF}.  However, there is still a significant mean-field contribution to the ratio up to $p_{miss}\approx 350$ MeV/c. Thus, the transition between these two regimes  occurs over the relatively narrow range of roughly 250 to 350~MeV/$c$.

\subsection{The Transition from Tensor to Central Correlation at High Momenta}
$np$ dominance holds for relative momenta from about $300-600$~MeV/$c$, the range driven by the tensor force and the minimum in the $pp$ $s$-wave momentum distribution. But what happens at higher relative momenta? At these momenta, the scalar (central) repulsive core of the $NN$ interaction becomes important. Since the central force does not distinguish among $np$, $pp$, and $nn$ pairs, we expect $pp$ and $nn$ pairs to become more common relative to $np$ pairs. This is sometimes called the transition from tensor to central correlations.

\begin{figure}[htb]
\begin{center}
\includegraphics[width=3.9in]{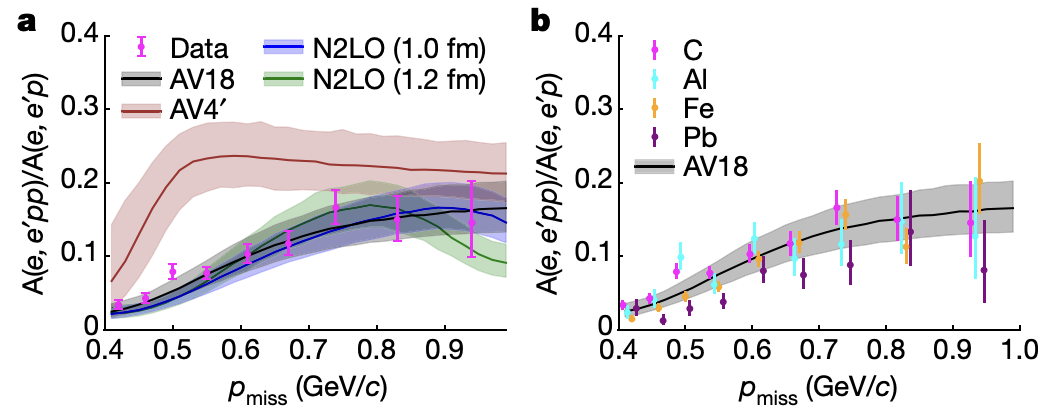}
\caption[]{\label{fig:ratio-schmnature} Event yield ratios of $(e,e'pp)/(e,e'p)$ plotted as a function of the $(e,e'p)$  missing momentum. Left, (a) shows measurements for carbon alongside theoretical predictions from the GCF framework employing different $NN$ interaction models. Right, (b) presents a comparison of C, Al, Fe, and Pb data with the GCF AV18 carbon calculation. Shaded band widths and data error bars indicate model systematic and statistical uncertainties, respectively, at the $1\sigma$ confidence level. Figure from~\cite{CLAS:2020mom}.}
\end{center}
\end{figure}

By studying carbon $(e,e'pp)$ and $(e,e'p)$ events as a function of missing momentum from 400 to 1000~MeV/$c$, \cite{CLAS:2020mom} found that the $pp$ to $p$ coincidence yield is suppressed at $p_{miss} = 400$~MeV/$c$ where $np$ tensor pairs dominate, and the yield rises to roughly 16\% by $p_{miss}\approx800$~MeV/$c$, which is deep into the
regime where the central (scalar) component of the $NN$ interaction has taken over, see Fig.~\ref{fig:ratio-schmnature}. This value is consistent with equal numbers of $pp$, $pn$, and $nn$ pairs (the central force limit, also called the ``scalar limit"), appropriately modified for the kinematics of the CLAS detector.

Crucially, this behavior was seen for calculations using many different $NN$ potentials — from the relatively hard AV18 potential to much softer chiral effective field theory (EFT) potentials (N2LO). Despite large differences in the absolute probability of SRC pair formation, all potentials including a tensor interaction predicted the same trend for the $pp/p$ ratio. The AV4' potential, which omits the tensor interaction, predicted a very different  ratio, inconsistent with the data.  

This transition was confirmed by \cite{Korover:2020lqf} by measuring $(e,e'pn)$ events and showing that the $(e,e'pn)/(e,e'p)$ ratio was consistent with unity (i.e., nearly every high-momentum proton has a neutron partner) for missing momenta from 300 to 1000~MeV/$c$, while the $(e,e'pp)/(e,e'p)$ ratio approached the scalar limit at the largest missing momenta.

\begin{figure}[htb]
\begin{center}
\includegraphics[width=2.5in]{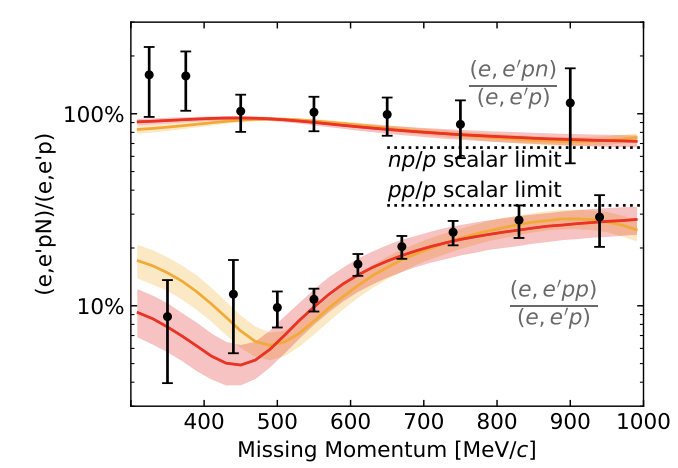}
\caption[]{\label{fig:korover-scalarlimit} C$(e,e'pp)/$C$(e,e'p)$ and C$(e,e'pn)/$C$(e,e'p)$ cross-section ratios compared with GCF calculations using the AV18 (dark band) and N2LO (light band) $NN$ interaction models. The dashed line marks the scalar limit from nucleon counting. Figure from~\cite{Korover:2020lqf}.}
\end{center}
\end{figure}

The tensor-to-scalar transition is not unique to heavier nuclei. In electron scattering on $^4$He at $Q^2= 2$~GeV$^2$ and $x_B > 1$, \cite{korover14} found that the fraction of two nucleon knockout events appearing in the $pp$ relative to the $pn$ channel climbs from about 5\% at 500~MeV/$c$ to roughly 12\% at 750 MeV/$c$, consistent with the observations from \cite{Korover:2020lqf,CLAS:2020mom}.

\subsection{Measuring the Center of Mass Motion of SRC Pairs}
\begin{figure}[htb]
\begin{center}
\includegraphics[width=2.5in]{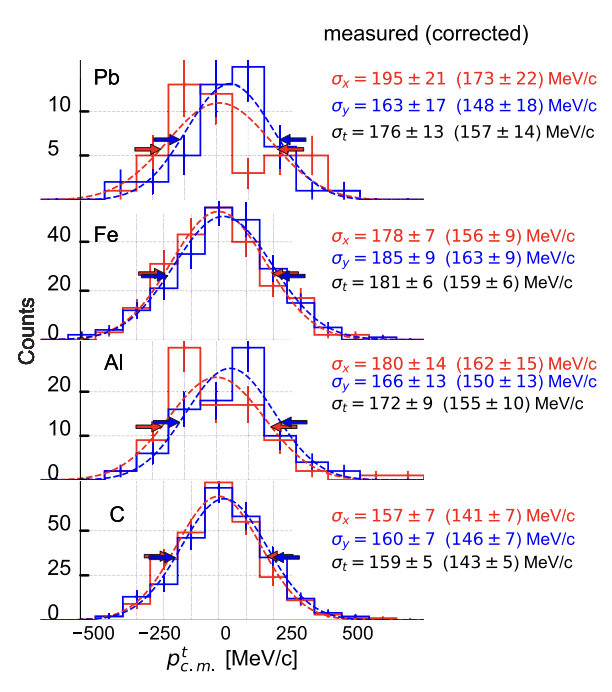}
\caption[]{\label{fig:pcm-clas} Distributions of $A(e,e'pp)$ events as a function of the perpendicular components of $\vec p_{cm}$ relative to $\vec p_{miss}$, with the $\hat{x}$ and $\hat{y}$ directions shown in red and blue, respectively. The distributions are displayed before CLAS acceptance corrections, and the dashed curves show Gaussian fits to the data. The fit values are obtained after correcting for detector acceptance. Reprinted figure with permission from \cite{Cohen:2018gzh}. Copyright 2018 by the American Physical Society.}
\end{center}
\end{figure}

The momentum-space definition of an SRC pair: large relative momentum $p_{rel}$ combined with smaller center of mass momentum $p_{cm}$ can be tested directly. \cite{Cohen:2018gzh} measured $p_{cm}$ from the total momentum of both outgoing protons in $(e,e'pp)$ events before the photon was absorbed ($\vec p_{cm} = \vec p_1 + \vec p_2 - \vec q$),  see Fig.~\ref{fig:pcm-clas}. They found that the $p_{cm}$ distribution in each transverse direction (perpendicular to the momentum transfer) was well described by a Gaussian with a standard deviation width of 140–170~MeV/$c$, see~\cite{Cohen:2018gzh}.

\begin{figure}[htb]
\begin{center}
\includegraphics[width=2.5in]{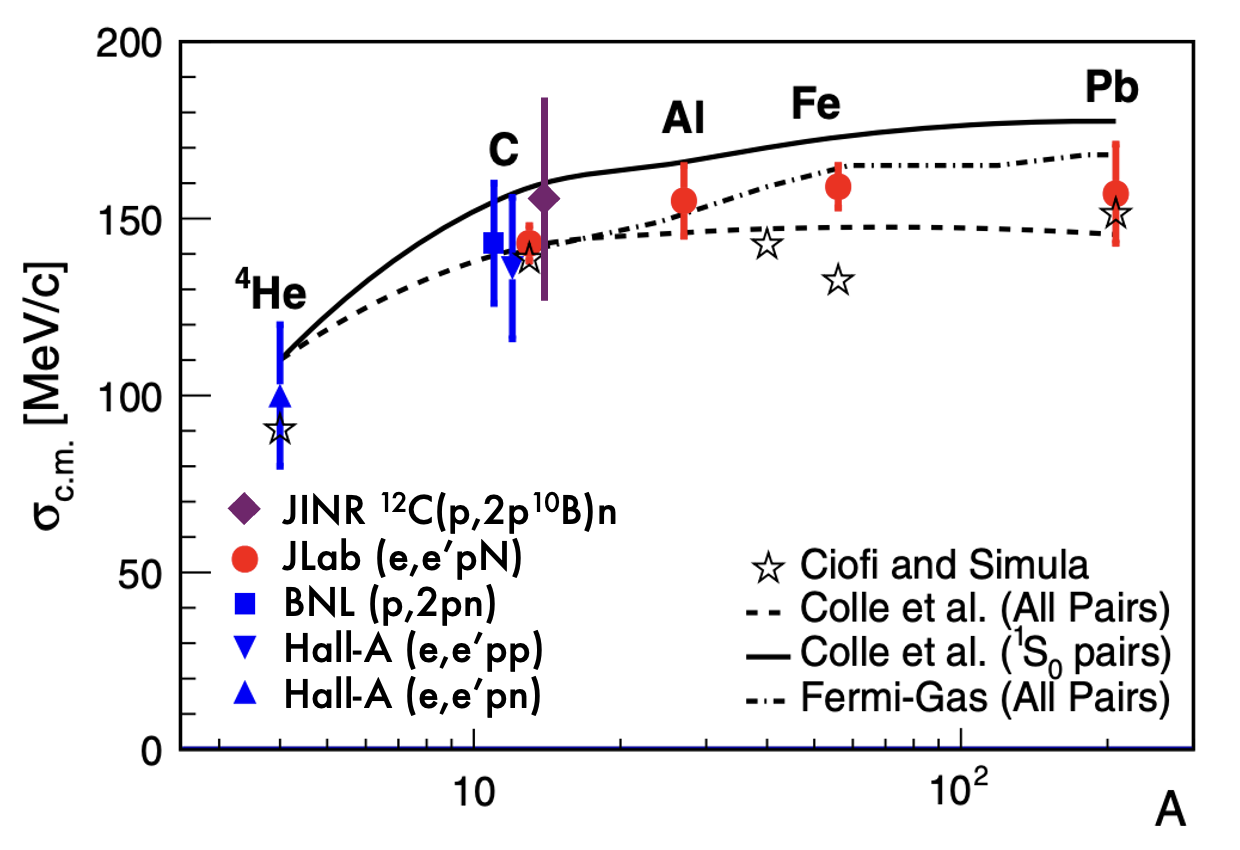}
\caption[]{\label{fig:pcm-jinr} Width of the one-dimensional center of mass momentum distribution as a function of nuclear mass. Blue circles show $(e,e'pN)$ results from~\cite{shneor07}, \cite{korover14}, and \cite{Cohen:2018gzh}; the red diamond shows $^{12}$C$(p, 2p^{10}$B) from \cite{Patsyuk:2021jea}; and the green square shows $(p,2pn)$ from \cite{tang03}. Theoretical curves from \cite{CiofidegliAtti:1995qe} (open stars), \cite{Colle:2013nna} for all mean-field pairs (dashed) and $^1S_0$ pairs only (solid), and a Fermi-gas combinatoric estimate using Fermi momenta from \cite{moniz71} are shown for comparison. Figure adapted with permission from \cite{Cohen:2018gzh} updated with data from \cite{Patsyuk:2021jea}. Copyright 2018 by the American Physical Society.}
\end{center}
\end{figure}

The center of mass momentum distribution was also measured in inverse kinematics by~\cite{Patsyuk:2021jea} in which the inverse-kinematics geometry enables the SRC-pair center of mass momentum to be obtained from the recoiling $A-2$ fragment in the projectile rest frame. The distributions agree well for the various nuclei. For $4$He, the width is about 100~MeV/$c$. For carbon, the width is about 140~MeV/$c$ and for heavier nuclei (Al, Fe, Pb), it is about 150~MeV/$c$. The combined results of the nuclear dependence of the measured $p_{cm}$ widths from is shown in Fig.~\ref{fig:pcm-jinr}. 

These widths are consistent with SRC pairs forming from two nucleons drawn randomly from the mean-field nuclear momentum distribution. This makes physical sense in that SRC pairs are not special pre-formed objects in the nucleus. They are fluctuations that arise dynamically when any two nucleons happen to approach each other closely, and their center of mass momentum reflects the typical momenta of mean-field nucleons.

\subsection{Inverse Kinematics}
A limitation of  electron scattering from fixed targets is the inability to measure the residual nucleus. The residual nucleus can be detected in either $eA$ collider measurements or in inverse kinematics, where a high-energy nucleus is scattered from a stationary proton target.  Detecting a specific residual nucleus can reduce contributions from other reaction mechanisms.

A first experiment scattered a  48~GeV/$c$ $^{12}$C ion beam from the Nuclotron accelerator at JINR (the Joint Institute for Nuclear Research in Russia) from a proton target, see~\cite{Patsyuk:2021jea}. The ion beam served as the nuclear target, and the target proton served as the probe. The Nuclotron $^{12}$C beam was triggered on hard $pp$ scattering at large angle and the data were sorted by the surviving spectator fragment, $^{10}$Be, $^{10}$B, or $^{11}$B, so that each event class corresponds to removing a specific $NN$ pair from $^{12}$C. The reaction ${12}$C($p,2p^{10}$B)$n$ happens when an $np$ SRC pair in the carbon projectile is struck (removing one proton and one neutron, leaving $^{10}$B), while ${12}$C($p,2p^{10}$Be)$p$ indicates a $pp$ SRC pair was struck.  \cite{Patsyuk:2021jea} found 23 events of the $pn$ type and only two of the $pp$ type,  consistent with $np$ dominance. 

\begin{figure}[htb]
\begin{center}
\includegraphics[width=3.5in]{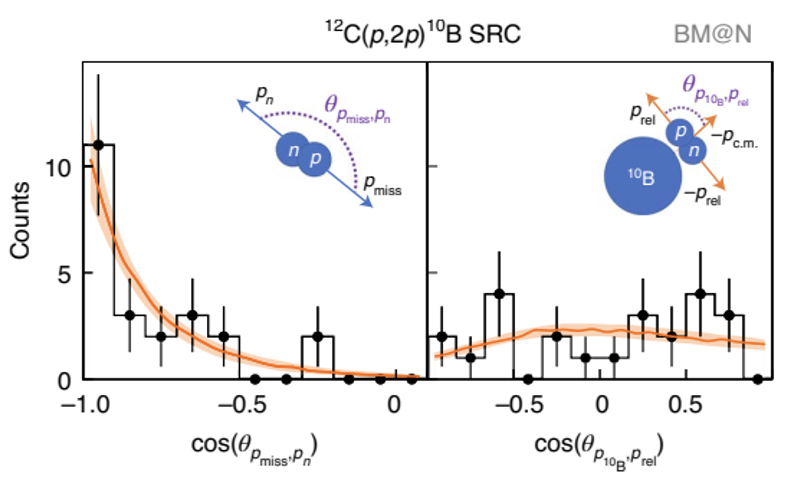}
\caption[]{\label{fig:jinr-angles} Left, the opening angle of the $np$ SRC pair; and  right, the opening angle between the pair relative  and center-of-mass momenta.  All angles are calculated in the rest frame of the incident $^{12}$C nucleus. The orange curves show the results of a GCF calculation assuming scattering from SRC pairs.  The width of the orange bands and the data error bars show the systematic uncertainties of the model and the data statistical uncertainties, respectively, each at the $1\sigma$ confidence level. Figure from~\cite{Patsyuk:2021jea}.}
\end{center}
\end{figure}

They also measured angular correlations consistent with scattering from nucleons in SRC pairs. The momentum of the $A-2$ nucleus equals the center-of-mass momentum of the SRC pair  (in the rest frame of the incident nucleus). They found that the $np$ pair was predominantly back-to-back, see Fig.~\ref{fig:jinr-angles}(left) and the angles of the center-of-mass pair momentum and the relative pair-momentum  were uncorrelated, as expected for SRC-pair knockout.  See Fig.~\ref{fig:jinr-angles}(right).  The consistency between electron- and proton-scattering experiments helps establish the probe independence of SRC physics.  

Future experiments will achieve higher statistics.  They will also be able to study SRC pairs in radioactive nuclei which are too unstable to be used as conventional targets.

\subsection{Photoproduction Reactions}

\begin{figure}[htb]
\begin{center}
\includegraphics[width=4in]{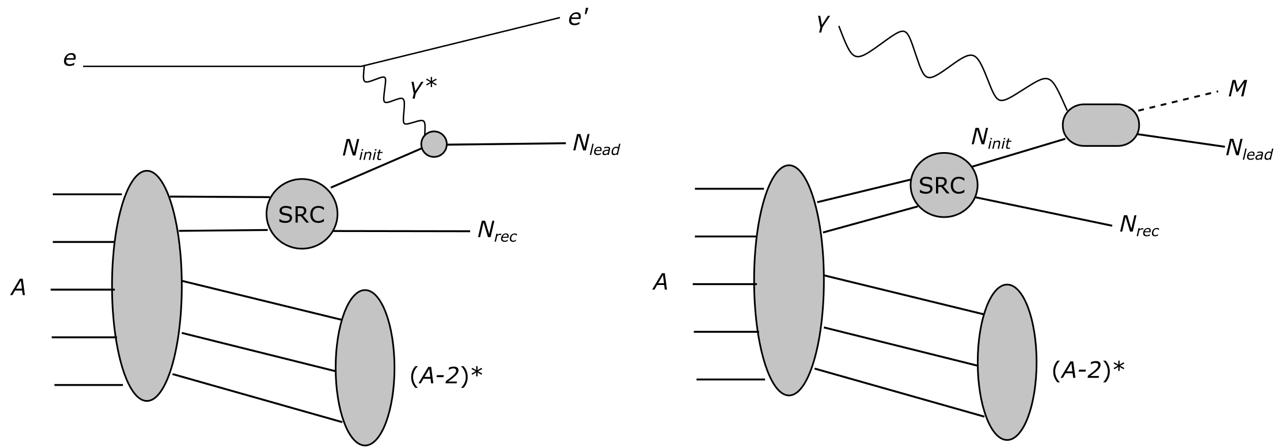}
\caption{\label{fig:src-rxn-photon} Left, an electron scatters quasielastically from a nucleon in a SRC pair.  Right, a real photon is absorbed on one nucleon in a SRC pair, producing a meson as described in the Vector-Meson Dominance model. In both reactions,  the correlated nucleon partner is also emitted and the  $A-2$ nucleus is intact. }
\end{center}
\end{figure}

We can also measure the effects of SRCs in nuclei through photon absorption experiments. A real photon beam probes the same nuclear ground state correlations through an entirely different kinematic and reaction pathway as compared to quasi-elastic electron scattering, see Fig~\ref{fig:src-rxn-photon}. Real photons have $\vert\vec q\thinspace\vert = \nu$ and thus $Q^2=0$.  This means that the dominant interaction will be quasi-elastic meson photoproduction on one nucleon of a correlated pair, with the second spectator nucleon also detected.   

If the SRC contact densities and pair fractions extracted from electron scattering are genuinely properties of the nuclear ground state, thus universal across probes at a given resolution, then they must appear consistently in photonuclear reactions. Evaluating SRCs using different probes would transform the SRC non-observable from a phenomenological construct to a scheme-aware quantity that can be evaluated across probes and kinematics.
Jefferson Lab's Hall D hosted the first such experiment to investigate SRCs using photonuclear reactions, see the experimental description in~\cite{hen2020studyingshortrangecorrelationsreal}. The setup directed tagged photons with energies from 6--10.6~GeV onto targets of deuterium, helium, and carbon. The GlueX spectrometer's large acceptance was used to reconstruct the multi-particle final states needed to isolate two-nucleon emission accompanied by meson photoproduction. Ongoing analysis of the data is exploring SRC processes involving $\rho^0$ and $\rho^-$ and seeks to demonstrate that the dataset is sensitive to the missing momentum observables connected to nuclear substructure. \cite{pybus2024measurementnearsubthresholdjpsi} observed the first $J/\Psi$ production from a nucleus, thus establishing the reaction and experiment could be sensitive to such observables. 

\section{Counting Short-Range Correlated Pairs with Inclusive Scattering \label{sec:a2}}
\subsection{Cross Section Ratios and Scaling}
\begin{figure}[htb]
\begin{center}
\includegraphics[width=2.5in]{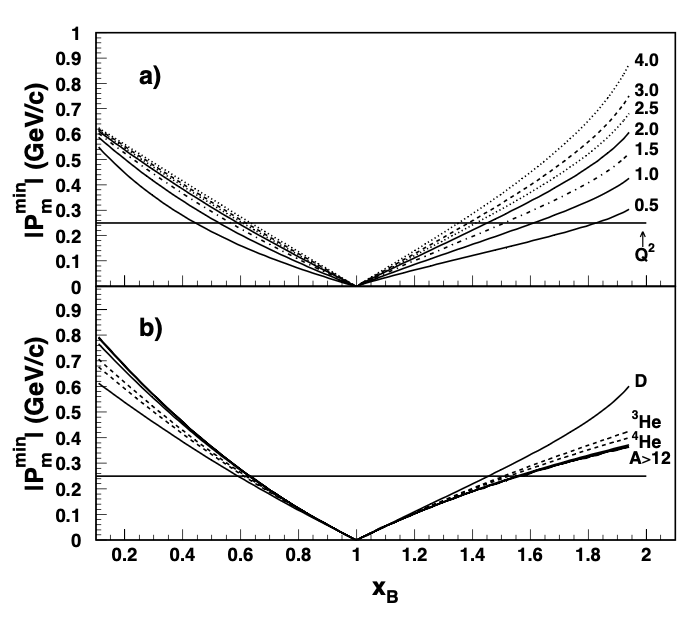}
\caption[]{\label{fig:pmin} The relationship between the minimum recoil momentum and $x_B$ a) for scattering from deuterium at different $Q^2$ and b) for scattering from different nuclei at $Q^2=2$ GeV$^2$. The horizontal lines are at 250 MeV/c, the Fermi momentum for heavy nuclei. Figure adapted from~\cite{egiyan03}.}
\end{center}
\end{figure}

We can also use inclusive quasi-elastic electron scattering to study nucleon motion in nuclei and specifically the high-momentum nucleons in SRC pairs.  Inclusive scattering at $x_B\ge 1$ can provide complementary information to that of more exclusive reactions, such as $(e,e'p), (e,e'pn)$ etc.  In quasielastic $A(e,e')$ scattering, $x_B, Q^2$ and the minimum $A-1$ recoil momentum are related by energy and momentum conservation (from \cite{egiyan03}):
\begin{equation}
    (q^\mu+p_A^\mu+p_{A-1}^\mu)^2 = (p_f^\mu)^2= m^2
    \end{equation}
where $q^\mu, p^\mu_A, p^\mu_{A-1}$ and $p_f^\mu$ are the four-momenta of the virtual photon, target nucleus, residual $A-1$ nucleus, and the knocked-out nucleon, respectively.  (only $q$ and $p_A$ are known).  This can be rearranged to give:
\[
  \Delta M^2-Q^2+\frac{Q^2}{m x_B}\left(M_A - \sqrt{M_{A-1}^2 + p_{miss}^2}\right)-2\vec q\cdot{\vec p}_{miss} - 2M_A \sqrt{M_{A-1}^2 + p_{miss}^2} = 0
\]
where $\Delta M^2 = M_A^2 - M_{A-1}^2 - m^2$ and $\vec p_{miss}= \vec p_f-\vec q = -\vec p_{A-1}$ is the recoil momentum.  This gives a relationship between the minimum recoil momentum $\vert\vec p_{miss}\vert$ and $x_B$ at fixed $Q^2$.  This kinematic relationship is shown in Fig.~\ref{fig:pmin}.

However, when an electron scatters from a nucleon in an SRC pair, the recoil momentum is carried predominantly by its correlated partner.  Thus, the appropriate minimum momentum for a given $Q^2$ and $x_B$ is that of the deuteron as shown  in the top half of Fig.~\ref{fig:pmin}.  This shows that the threshold for electron scattering from nucleons in SRC pairs occurs at about $Q^2=1.5$ GeV$^2$ and $x_B=1.5$.

\begin{figure}[htb]
\begin{center}
\includegraphics[width=2.5in]{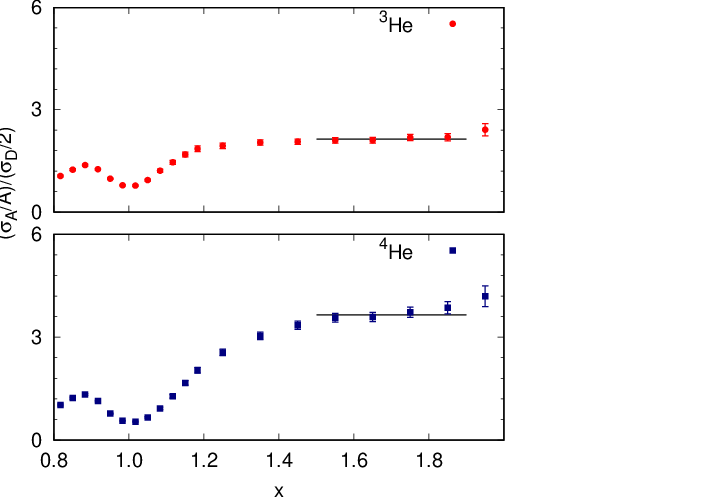}
\includegraphics[width=2.5in]{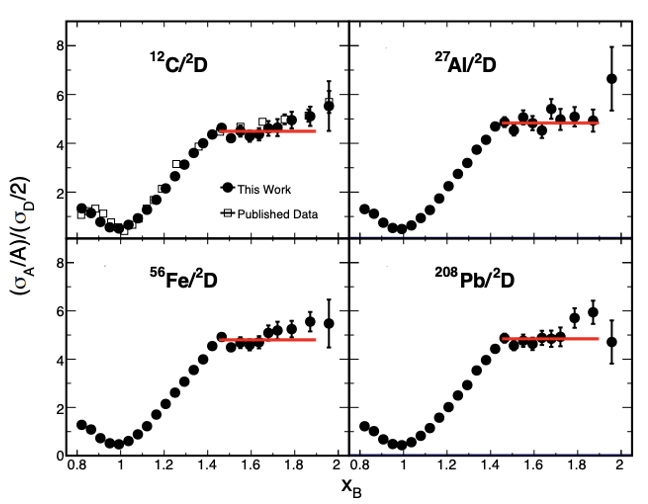}
\caption[]{\label{fig:a2-scaling} Per-nucleon cross-section ratios of nucleus $A$ to deuterium in quasielastic kinematics ($0.8\le x_B\le 1.9)$. a) $^3$He and $^4$He are shown in the left figure adapted with permission from \cite{Fomin:2012}. Copyright 2012 by the American Physical Society and b) C, Al, Fe and Pb,  solid points are from \cite{Schmookler:2019nvf}, open squares are from~\cite{Fomin:2012}. Red lines show constant fits in the quasielastic region where scaling occurs. Error bars include statistical and point-to-point systematic uncertainties at $1\sigma$, and no isoscalar correction has been applied. Right figure is from~\cite{Schmookler:2019nvf}.}
\end{center}
\end{figure}

Since the momentum distributions of SRC pairs are nucleus independent, we expect that the cross-section ratios for inclusive electron scattering from nucleus $A$ to deuterium at fixed $Q^2$ should be independent of $x_B$ for $Q^2\ge 1.5$ GeV$^2$ and $1.5 \le x_B\le 2$, see~\cite{Frankfurt:1993sp}. This scaling behavior was subsequently observed in measurements at SLAC (\cite{Frankfurt:1993sp}) and at Jefferson Lab (\cite{egiyan03,egiyan06,Fomin:2012,Schmookler:2019nvf}). Scaling sets in above an initial-momentum threshold of $\approx275\pm25$~MeV/$c$, slightly larger than the nuclear Fermi momentum. The scaling plateau means that the shapes of the high-momentum nucleon distributions in all nuclei are the same above 275~MeV/$c$, but they differ in overall magnitude. The height of the plateau, or magnitude $a_2$, gives the relative probability of finding a high-momentum nucleon in nucleus $A$ compared to the deuteron.  

The measured plateau values, $a_2$, increase systematically from light to heavy nuclei. $^3$He has a value of two, approximately twice as many high-momentum nucleons per nucleon as deuterium. $^4$He has a value of 3.6, and $^{12}$C has a value of 4.5, see~\cite{Fomin:2012,hen11}. Heavier nuclei such as Al, Fe, Au tend to saturate around a value of 4.8, see \cite{Schmookler:2019nvf} and Fig.~\ref{fig:a2-scaling}. Two systematic effects must be folded in before these ratios can be read as true pair counts: the residual $A-2$ system is generally left in an excited state after the pair removal, and the SRC pair itself carries a non-zero center of mass motion that smears the missing momentum distribution, see~\cite{Weiss:2020bkp}.

\subsection{Three Nucleon and Higher Order Correlations}
The SRC pairs we have discussed are two-nucleon correlations. This leads one to wonder about the possibility of three-nucleon short-range correlations (3N-SRCs), in which three nucleons are close together? Theoreticians expect three-nucleons SRCs to contribute at even higher momenta above about 600~MeV/$c$ and at shorter inter-nucleon distances ($r<0.5$~fm), capable of probing the repulsive core of the $NN$ interaction.

The observation of the scaling plateau for $2<x_B<3$ in inclusive electron scattering would indicate three-nucleon correlations. No such scaling plateau has been conclusively observed yet, see~\cite{Egiyan:2006,Ye:2017mvo,ZHANG2025140087,Higinbotham:2014xna}.  Further measurements using $\alpha_{2N}$ and $\alpha_{3N}$ scaling variables~ as in~\cite{Day:2018nja,Fomin:2017ydn}, especially on light nuclei, show hints of $3N$ correlations shown in~\cite{Li:2024rzf}, but more data is needed.
\begin{figure}[htb]
\begin{center}
\includegraphics[width=2.5in]{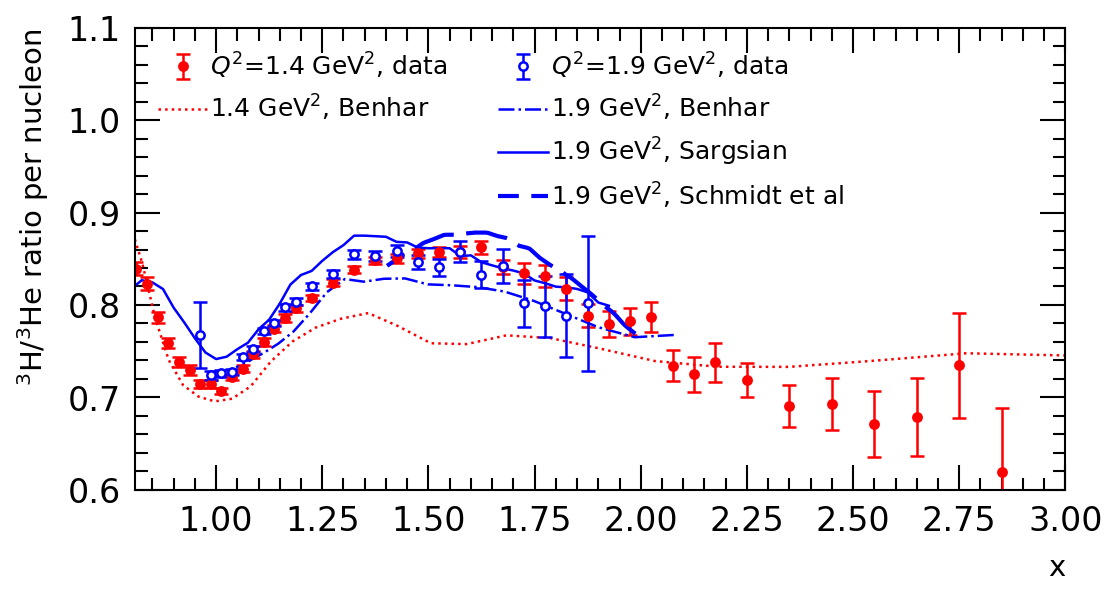}
\caption[]{\label{fig:A3-ratio} 
Measured ratio of the inclusive tritium to helium-3 $(e,e')$ cross sections, $\sigma_{3H}/\sigma_{3He}$, as a function of $x_B$ with the data from~\cite{Li:2022fhh} and reproduced from~\cite{Li:2024rzf}. Data are shown at $Q^2\approx 1.4$~GeV$^2$ (red) and $Q^2\approx 1.9$~GeV$^2$ (blue), together with theoretical calculations from~\cite{sargsian14}, \cite{Benhar:1993ja,Benhar:2013dq}, and \cite{Schmidt:2024fok}. Figure courtesy S.Li.}
\end{center}
\end{figure}
The cross-section ratio of $^3$H to $^3$He as measured in the kinematic region $x_B>1$ is shown in Fig.~\ref{fig:A3-ratio} with results compared against multiple theoretical predictions. Within the SRC-dominated regime spanning $1.4 < x_B < 1.7$, the measured ratio averages $0.854 \pm 0.010$. A factorized cross section approach reproduces this value well when employing a spectral function obtained from exact three body ground state calculations that incorporate spectator final state interactions while omitting irreducible three-body forces, see~\cite{wiringa95,CiofidegliAtti:2004jg}. Additional theoretical estimates have been provided at $Q^2\approx1.9 $~GeV$^2$ (\cite{sargsian14}), as well as at $Q^2\approx1.4$ and 1.9~GeV$^2$ (\cite{Benhar:1993ja,Benhar:2013dq}). The convergence of these independently developed calculations with experimental data reflects a remarkable understanding for few-body nuclear theory.

\section{SRCs and nucleon structure}

\begin{figure}[htb]
\begin{center}
\includegraphics[width=6in]{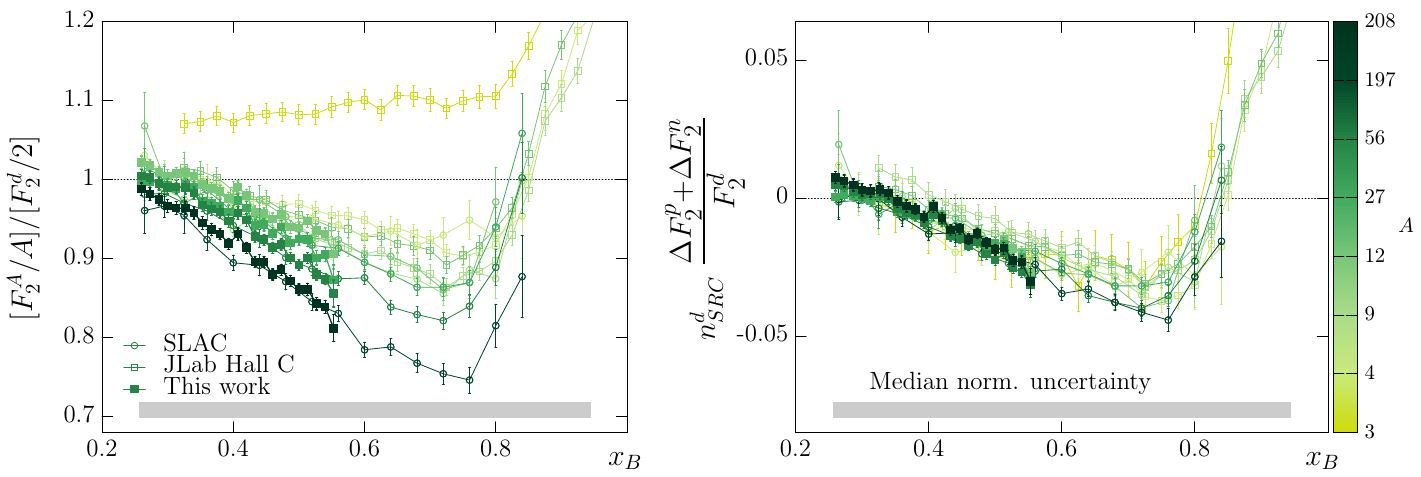}
\caption{\label{fig:emc_scaling} (left) Per-nucleon inclusive DIS cross-section ratios of nucleus $A$ to deuterium (the EMC Effect) plotted vs. $x_B$ and (right) the modification of SRC pairs, as described by Eq.~\ref{eq:EmcUniversal}.  Solid points are from \cite{Schmookler:2019nvf}, open circles are from ~\cite{Gomez94}, open squares are from~\cite{Seely:2009gt}.Error bars include statistical and point-to-point systematic uncertainties at $1\sigma$, and no isoscalar correction has been applied. The gray bands show the $1\sigma$ median normalization uncertainty.  Different colors correspond to different nuclei, as shown by the color scale on the right.  Figure adapted from~\cite{Schmookler:2019nvf}.}
\end{center}
\end{figure}

One of the biggest open questions in nuclear physics is how the partonic (quark and gluon) structure of nucleons is modified when they are inside the nucleus. The longitudinal momentum distribution of quarks in nucleons and nuclei is measured by inclusive $A(e,e')$ deep inelastic scattering (see the chapters in this volume on ``Deeply inelastic scattering (DIS) experiments – overview'' and ``the EMC Effect'').  The per-nucleon inclusive DIS cross-section ratio of nucleus $A$ to $d$ is approximately unity at $x_B=0.3$, decreases linearly to $x_B\approx 0.75$ and then increases rapidly due to the effects of nucleon motion (see Fig.~\ref{fig:emc_scaling} left).   The  linear decrease for $0.3\le x_B\le 0.75$  is referred to as the EMC effect.  This effect cannot be described by theoretical models without including the effects of nucleon modification, see~\cite{Hen:2016kwk}. However, so many different models can describe the EMC Effect that one physicist claimed that EMC stands for ``Every Model is Cool'' (\cite{millerPC}).

\begin{figure}[htb]
\begin{center}
\includegraphics[width=4in]{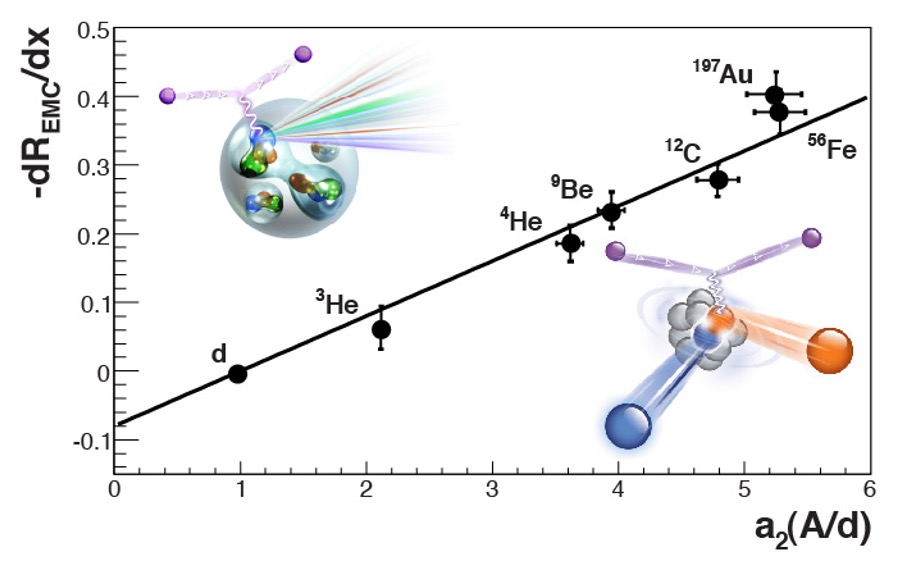}
\caption{\label{fig:emcsrc} (left) The slope of the isoscalar-corrected EMC Effect ratio for $0.35\le x_B\le 0.7$ plotted versus the isoscalar-corrected measured $a_2$ ratios for $x_B\ge 1.4$.  The data is from~\cite{weinstein11,Hen12,Hen:2016kwk}.  The straight line is a linear fit to the data, constrained to pass thru the deuteron point at (1,0). Figure modified from~\cite{Hen:2016kwk}.}
\end{center}
\end{figure}

One important clue to the origin of the EMC Effect is the linear correlation between  the magnitude of the EMC effect (as measured by its slope) and the relative fraction of high-momentum nucleons in nuclei (as measured by the inclusive QE cross section ratios $a_2$ described in Section~\ref{sec:a2}), see Refs.~\cite{weinstein11,Hen12,Hen:2016kwk}. As shown in Fig.~\ref{fig:emcsrc}, this correlation strongly implies that the EMC Effect is due to the strong modification of nucleons in SRC pairs caused by the strong forces between them.  One opposing theory attributes the EMC Effect to a smaller modification of all nucleons, caused by the average nucleon-nucleus interaction~\cite{Cloet:2006bq,Cloet:2005rt}.

If we assume that only the nucleons in SRC pairs are modified in the nucleus, then we can write the nuclear structure function $F_2^A$ as 
\begin{equation}
F_2^A=ZF_2^p+NF_2^n +n^A_{SRC}(\Delta F_2^p+\Delta F_2^n)
\end{equation}
where $F_2^p$ and $F_2^n$ are the free proton and neutron structure functions, $\Delta F_2^p$ and $\Delta F_2^n$ are the differences in the bound structure functions of nucleons in SRC pairs, and $n^A_{SRC}$ is the number of $np$-SRC pairs in nucleus $A$.  Dividing this equation by the corresponding equation for the deuteron, rewriting $F_2^n=F_2^d-F_2^p-n^d_{SRC}(\Delta F_2^p+\Delta F_2^n)$, and rearranging, we get:
\begin{equation}
    \frac{n^d_{SRC}(\Delta F_2^p+\Delta F_2^n)}{F_2^d} = \frac{F_2^A/F_2^d - (Z-N)F_2^p/F_2^d - N}{(A/2)a_2-N}
    \label{eq:EmcUniversal}
\end{equation}
where  all the quantities on right side are experimentally measured.  The structure-function modification of nucleons in SRC pairs, $(\Delta F_2^p+\Delta F_2^n)$ is nucleus independent, see Fig.~\ref{fig:emc_scaling} right.  All of the curves for nuclei from $A=3$ to 208 have the same slope (even if the experimental normalizations might vary), indicating that a universal modification of $F_2$ for nucleons in SRC pairs can describe the EMC effect.

\begin{figure}[htb]
\begin{center}
\includegraphics[width=3in]{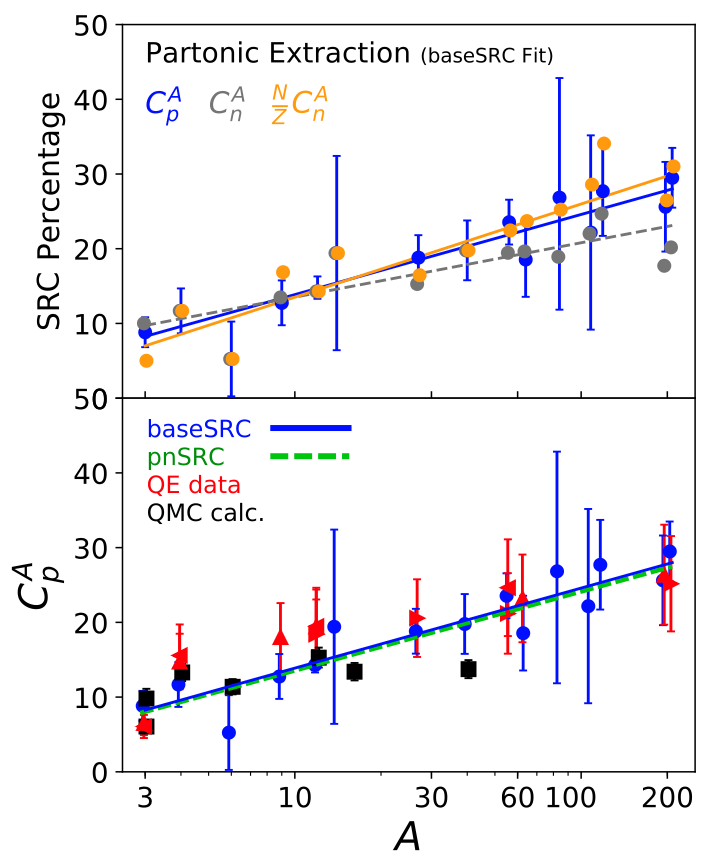}
\caption{\label{fig:ncteq} (top) Comparison of the nuclear structure parameters $C^A_p, C^A_n, $ and $(N/Z)C^A_n$.  The lines represent logarithmic fits to the corresponding quantities; (bottom) the comparison of fit $C^A_p$ values and the SRC abundances extracted from quasielastic data (\cite{Egiyan:2006,Fomin:2012,Schmookler:2019nvf}) and from Quantum Monte Carlo (QMC) calculations (\cite{Cruz-Torres:2019fum}). Figure from~\cite{PhysRevLett.133.152502}.}
\end{center}
\end{figure}

A global analysis of high-energy data from lepton deep-inelastic scattering, Drell-Yan, and $W$ and $Z$ boson production simultaneously extracted the universal effective distribution of quarks and gluons inside correlated nucleon pairs, and their nucleus-specific fractions, see~\cite{PhysRevLett.133.152502}.  They extended the QCD Parton Model analysis using a factorized nuclear structure model incorporating both individual nucleons and pairs of correlated nucleons.  This allowed them to separately fit the fraction of nucleons in SRC pairs in each nucleus, and the PDFs  for nucleons inside SRC pairs, see Fig.~\ref{fig:ncteq}top.  The extracted SRC proton probabilities are in good agreement with both data and calculation, see Fig.~\ref{fig:ncteq}bottom.  This is remarkable, because the fit did not use any of the quasielastic scattering data previously used to determine SRC probabilities in nuclei.  It is more evidence that SRCs are scale and scheme independent.  The fact that the obtained fractions of SRC pairs agree with their previous extractions from the low-energy quasielastic data establishes a direct link between high-energy partonic properties and lower-energy nuclear physics.

\section{Theoretical approaches}

A precise calculation of an SRC-related reaction cross section would calculate the unfactorized cross section from  the matrix elements of the appropriate operator sandwiched  between the initial and final states, averaged over all possible initial states and integrated over all possible final states.  However, while nuclear initial states are becoming calculable, due to their high energies,  final states are not calculable.

Instead, almost all calculations of SRC-related reaction cross sections assume factorization of the cross section into a reaction part and a nuclear structure part: 
\begin{equation}
    \dfrac{d\sigma}{d\nu d\Omega_e dE_{miss} d\Omega_p} = K\sigma_{ep}S(E_{miss},p_{miss})
\end{equation}
as discussed in Section~\ref{sec:srckin}.
The attenuation of the outgoing nucleon(s) due to final state interactions  is typically treated separately using the Glauber approximation.  



This section describes the various approaches  for calculating the nuclear initial states. These range from exact calculations for light nuclei to approximate methods for heavier nuclei.

\subsection{The High-Energy Perspective: Frankfurt and Strikman}\label{sec:theory:fs}
The modern foundational theoretical treatment of SRCs, encompassing QCD and relativistic quantum field theory, was established by~\cite{Frankfurt81}. Their work predicted experimental SRC signatures in inclusive scattering experiments and remains relevant to all SRC interpretations. 

When studying the dynamics inside a nucleus, it is most natural to use the light cone formalism in which the nuclear state is described in terms of the constituent nucleon momenta along the direction of the momentum transfer. In this way, a nucleus with mass $A$ can be described as a collection of nucleons with some that may, at any instant, form correlated clusters at short inter-nucleon separations. 

Their significant insight is that at large nucleon momenta, the dominant contribution to the nuclear wave function comes from the configurations of two nucleons at short-range, thus forming the two-nucleon SRC. The remaining $A-2$ nucleons of the nucleus act as ``spectators" remaining near their mean field configurations. The two body dominance in this high momentum regime leads to specific experimental predictions. 

They predicted the existence of the plateaus in the high-$Q^2$ inclusive-scattering cross-section ratios of nucleus $A$ to deuterium at $1.5\le x_B\le 2$ (see Section~\ref{sec:a2}):
\begin{equation}
    \frac{\sigma_A / A}{\sigma_d / 2}
    \;\xrightarrow{\;1 < x_B < 2\;}\; a_2(A).
    \label{eq:scaling}
\end{equation}
$a_2(A)$ is the SRC scaling factor and is related to the probability of finding a proton-neutron pair in the correlated
$^3S_1$-$^3D_1$ state characteristic of the deuteron, and therefore it measures the ``number of deuteron-like pairs" per nucleon in nucleus $A$.

A direct consequence of this scaling is that the high-momentum tail of the nuclear single-nucleon momentum distribution $n_A(k)$ should have the same shape for all nuclei and be proportional to the deuteron
momentum distribution $n_d(k)$ for $k \gg k_F$:
\begin{equation}
    n_A(k) \;\approx\; a_2(A)\cdot n_d(k), \qquad k \gg k_F.
    \label{eq:hmt}
\end{equation}
This predicts a second type of factorization, where the nuclear momentum distribution can be approximated as a single-nucleon mean-field part at low momentum plus an SRC part at higher momentum. 

This framework extends to exclusive knockout reactions and to the phenomenological descriptions using nuclear contacts as universal coefficients that quantify SRC pair densities and connect the many-body nuclear wave function to the short-distance $NN$ physics (see Sect.~\ref{sect:GCF}).

\subsection{Ab Initio Quantum Monte Carlo Methods} \label{sec:theory:qmc}

For light nuclei, exact nuclear ground states are available from precise ab-initio calculations using variational methods to solve the Schrodinger equation for different $NN$ potentials.
The leading approach uses Variational Monte Carlo (VMC) and Green's Function Monte Carlo (GFMC), developed by~\cite{Carlson:2014vla,schiavilla07,wiringa14}.

The nuclear Hamiltonian is:
\begin{equation}
    H = \sum_i T_i + \sum_{i<j} v_{ij} + \sum_{i<j<k} V_{ijk}
    \label{eq:hamiltonian}
\end{equation}
where $T_i$ is the kinetic energy of nucleon $i$, $v_{ij}$ is the $NN$ potential (e.g., the Argonne $v_{18}$ (AV18)
two-body potential from~\cite{Wiringa:1994wb}), and $V_{ijk}$ is the three-nucleon potential (e.g., the Illinois-7 (IL7) three-body force). AV18 is fitted to the complete $NN$ scattering database and contains 18 operator
structures:
\begin{equation}
    v_{ij} = \sum_{p=1}^{18} v^p(r_{ij})\,O^p_{ij},
\end{equation}
where the operators $O^p_{ij}$ include central, spin-spin, isospin-isospin, tensor, spin-orbit, and charge-symmetry-breaking terms. The tensor component is especially important as it generates the $D$-state admixture in the deuteron and is the primary driver of SRCs in $T=0$ ($pn$) pairs.

In Variational Monte Carlo (VMC), a trial wave function $|\Psi_T\rangle$ is built by applying two- and three-body correlation operators to an antisymmetrized shell-model state and minimizing the energy expectation value via Monte Carlo integration over the full $3A$-dimensional configuration space. Green's Field Monte Carlo (GFMC) then projects out the exact ground state by imaginary-time evolution:
\begin{equation}
    |\Psi_0\rangle \propto \lim_{\tau\to\infty} e^{-(H-E_T)\tau}\,|\Psi_T\rangle,
    \label{eq:gfmc}
\end{equation}
yielding essentially exact energies and wave functions for $A \leq 12$.

These calculations directly confirm the SRC picture. The computed nuclear momentum distributions exhibit a universal high-momentum tail of the form of Eq.~\ref{eq:hmt}, and two-body density matrices in momentum space reveal that back-to-back pairs with
$k > k_F$ are overwhelmingly proton-neutron ($pn$), with $pp$ and $nn$ pairs suppressed by roughly an order of magnitude (\cite{schiavilla07}), see Fig.~\ref{fig:anl-NN-calc}. This $pn$ dominance is a direct consequence of the tensor force acting preferentially in the $T=0$ channel.  GFMC has also been extended to compute $(e,e')$ cross sections for electron scattering by evaluating Euclidean response functions via imaginary-time correlation functions, and recently the QMC spectral function has been used to calculate $(e,e'p)$ and $(e,e'2N)$ cross
sections, see Refs.~\cite{Carlson:2014vla,Lovato:2013cua}.

\subsection{The Similarity Renormalization Group}
\label{sec:theory:srg}

Calculating exact wave functions for heavy nuclei with hard potentials ($NN$ potentials with a very repulsive hard core that create high-momentum nucleons) is computationally impractical.  
The Similarity Renormalization Group (SRG) of~\cite{bogner07,Bogner:2012zm} uses unitary transformations to effectively ``soften" the potentials and dramatically reduce the high-momentum components of the nuclear wave function.  In order to calculate the same physics, the interaction operators need to be similarly transformed, see~\cite{Tropiano:2021qgf,Tropiano:2024bmu}.  

At high renormalization group (RG) resolution, SRCs are identified as high-momentum components in the nuclear wave function (as was discussed in most of this paper).  This results in scale separation where the nuclear wave function can be written as the sum of a single-nucleon mean-field part plus a correlated part. At lower RG resolution, SRC physics shifts from the nuclear wave function  to the more complicated reaction operators without changing the measured observables.

A continuous unitary transformation
$H_s = U_s H U_s^\dagger$ is applied to the nuclear Hamiltonian, with the flow equation:
\begin{equation}
    \frac{dH_s}{ds} = [\eta_s,\,H_s], \qquad \eta_s = [T_{\rm rel},\,H_s]
    \label{eq:srg}
\end{equation}
where $s$ parameterizes the flow (equivalently, a momentum resolution scale $\lambda_{\rm SRG} \sim s^{-1/4}$). As $\lambda_{\rm SRG}$ decreases, $H_s$ becomes band-diagonal in momentum space: matrix elements coupling low-momentum and high-momentum states are suppressed, and the nuclear wave function in the evolved basis contains far less high-momentum content.

The transformation is exactly unitary: all physical observables are unchanged provided that every operator is evolved consistently with the Hamiltonian. This has a profound implication for SRC physics predicting that the high-momentum tail of $n(k)$ is scheme-dependent. At high resolution ($\lambda_{\rm SRG} \sim 2\,{\rm fm}^{-1}$, close to the original hard AV18 potential), SRCs appear explicitly in the wave function; at low resolution ($\lambda_{\rm SRG} \sim 1\,{\rm fm}^{-1}$), they are ``hidden" in evolved two-body operators while the soft wave function appears uncorrelated. The SRG thus provides a rigorous language for what ``resolution" means in nuclear physics. 

Comparing SRC observables extracted at different scales, or between SRG-evolved and unevolved interactions with consistently evolved currents, is an important test of theoretical
consistency and a probe of the scale-independence of physical predictions, see~\cite{Bogner:2012zm}.

\subsection{Factorized Approaches}
\label{sec:theory:factorized}

While ab-initio calculations are exact for light nuclei (\cite{Carlson:2014vla}), approximate factorized approaches are essential for medium-to-heavy nuclei.   These approaches rely on scale separation to write the nuclear momentum distribution as the sum of a mean field part at low momentum plus a SRC part at high momentum. This is different from the cross section factorization mentioned above.

\subsubsection{Correlated Basis Functions and the Spectral Function Approach}\label{sec:theory:cbf}

\cite{CiofidegliAtti:1995qe} developed a detailed microscopic account of SRCs using Jastrow-correlated many-body wave functions:
$|\Psi_0\rangle = \mathcal{F}|\Phi_0\rangle$, where $|\Phi_0\rangle$ is the IPSM Slater
determinant and $\mathcal{F} = \prod_{i<j}f(r_{ij})$ is a product of pair correlation functions that suppress the wave function at short separations. A cluster expansion
evaluates expectation values systematically in increasing numbers of correlated particles.
 
The key result is the factorization of the momentum distribution at high $k$:
\begin{equation}
    n(k) \approx C_{NN}\cdot|\tilde{\phi}(k)|^2, \qquad k \gg k_F
    \label{eq:nkfactorized}
\end{equation}
where $\tilde{\phi}(k)$ is the Fourier transform of the correlated-pair relative-motion wave function and $C_{NN}$ is the nuclear contact for the relevant spin-isospin channel. The shape of the tail is universal (set by the $NN$ interaction); only the
amplitude $C_{NN} \propto a_2(A)$ varies with the nucleus. They also find that the SRC pairs are almost entirely due to the $NN$ tensor interaction, see~\cite{alvioli08}.

A related approach, the Low-order Correlation Operator Approximation (LCA) of Refs.~\cite{Vanhalst:2011es,Vanhalst:2012ur,Colle:2015ena,Ryckebusch:2019oya}, implements this correlation structure in coordinate space using a minimal set of three correlation functions (central, tensor, and spin-isospin) acting on mean-field wave functions, providing nucleus-independent predictions that extend naturally to heavy nuclei and can be systematically compared with VMC results.


\subsubsection{The Generalized Contact Formalism}\label{sect:GCF}
\textbf{Scale Separation and Universality:} The physics of SRC pairs operates at a scale of $\approx1$~fm or less, while the mean-field nuclear structure operates at scales of several fm, see~\cite{Miller:2020eyc}. When these two scales are well-separated, we can factorize the two-nucleon density in the nucleus into a nucleus-dependent part (encoding how often pairs approach each other) and a universal part (encoding what happens when they do).

\begin{figure}[htb]
\begin{center}
\includegraphics[width=4in]{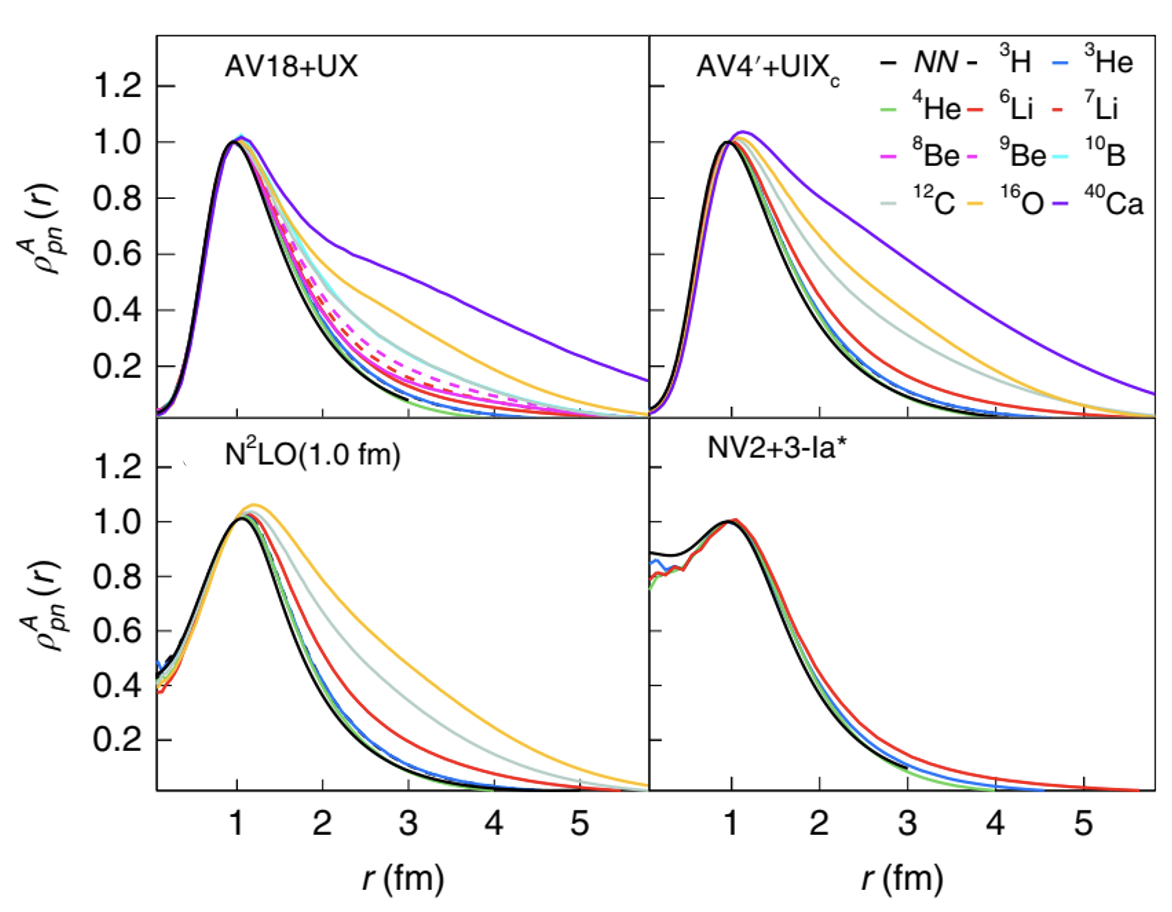}
\caption{\label{fig:gcf-universality} The center-of-mass-integrated $np$-pair  density plotted versus relative position $r$ for multiple nuclei plus the universal two-body function $|\phi_{np}^{s=1}|^2$ (black lines) for each of four $NN$ plus $3N$ potentials, the phenomenological Argonne AV18 + Urbana X (top left) and AV4' + Urbana IXc (top right) potentials, and the chiral N$^2$LO(1.0~fm) (bottom left) and NV2+3-Ia* (bottom right) potentials.  Each panel is normalized at $\sim$1~fm,  revealing that all nuclei share the same short-distance pair structure regardless of mass. Differences across the four panels reflect the sensitivity of many-body calculations to the choice of interaction, highlighting the inherent model dependence of these results. Figure from~\cite{Cruz-Torres:2019fum}.}
\end{center}
\end{figure}


Calculations confirm universality in that the two-nucleon position and momentum distributions for a variety of nuclei from helium to calcium and for different $NN$ potentials show that at short separation, or high relative momentum ($p_{rel}\geq k_F$), the pair density has the same shape in all nuclei, and only its overall magnitude varies.


The Generalized Contact Formalism (GCF) decribed by Refs.~\cite{Weiss:2015mba,Weiss:2016obx,Weiss:2018tbu,Cruz-Torres:2019fum} factorizes the two-body density at short range into a universal pair structure and a nucleus-specific normalization,
\begin{equation}
    \rho_A^{NN,\alpha}(r)=C_A^{NN,\alpha}\times|\psi_{NN}^{\alpha}(r)|^2
    \label{eq:gcf_pos}
\end{equation}
\begin{equation}
    \rho_A^{NN,\alpha}(q)=C_A^{NN,\alpha}\times|\psi_{NN}^{\alpha}(q)|^2
\end{equation}
Here $\psi_{NN}^{\alpha}$ is the universal short range pair wave function (independent of the nucleus) for an $NN$ pair of total spin $\alpha$ of 0 or 1, and the label $NN$ denotes $pp, pn,$ or $nn$, while the nuclear contacts $C_A^{NN,\alpha}$ carry all of the $A$-dependence. $r$ and $q$ are the relative position and momentum.  These contacts can be fit to theory calculations or to data.

Figure~\ref{fig:gcf-universality} shows the two-nucleon relative-position densities for different nuclei for four different $NN$ potentials calculated using quantum monte carlo techniques (see Section~\ref{sec:theory:qmc}).  The long-distance densities vary dramatically for the different potentials and different nuclei.  However, the short-distance densities are the same for all nuclei, even though they differ for the different potentials.  Thus, the $|\psi_{NN}^{\alpha}(r)|^2$ terms in Eq.~\ref{eq:gcf_pos} are identical for all nuclei for $r\le 1$ fm, see~\cite{Cruz-Torres:2019fum}.

\textbf{Contacts Confirm SRC Universality: }The contacts encode which pairs form most often and in which nuclei. The $pn$ contacts normalized per nucleon pair are relatively constant across all the measured nuclei, thereby confirming the universality of SRC formation. The spin-0 $pp$ contacts show a similar universality. This means that the probability of SRC pair formation per nucleon pair is roughly the same in all nuclei.

An important conclusion of the GCF analysis is that the formation of SRC pairs is a long-range (mean-field) phenomenon. While the properties of the pairs themselves are determined by the short-range interaction, the probability of two nucleons coming close enough to form a pair is determined by the long-range nuclear wave function. The contacts are therefore the interface between nuclear structure physics (which determines $C$) and few body physics (which determines $\psi$).

\textbf{Contact Ratios are Scheme-Independent: } While the individual contacts $C$ depend on the choice of $NN$ potential, the ratios of contacts for different nuclei are stable. 
\begin{figure}[htb]
\begin{center}
\includegraphics[width=2.5in]{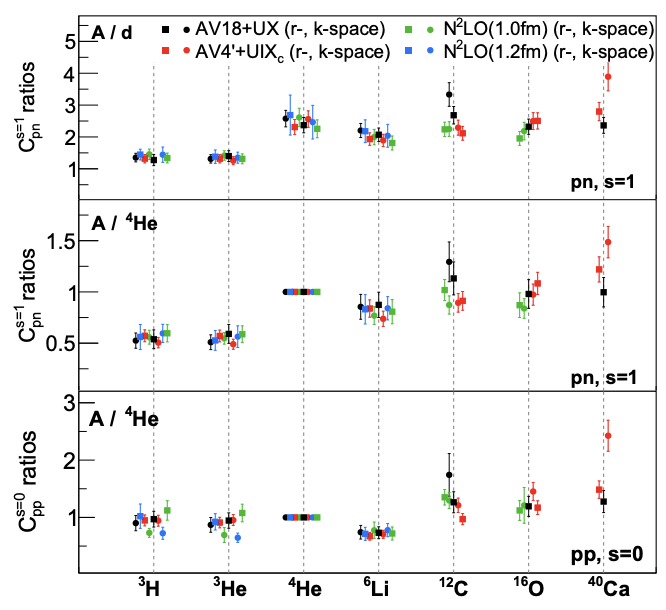}
\caption[]{\label{fig:nucl-contacts} Ratios of nuclear contact terms relative to deuterium (top) or $^4$He (middle and bottom), shown for spin-1 $pn$ pairs (top and middle) and spin-0 $pp$ pairs (bottom) across several nuclei. Contact term ratios were extracted using various NN+3N potentials in both coordinate space (squares) and momentum space (circles). For $^3$H in the spin-0 panel, the contact value corresponds to $C^{s=0}_nn$, as no pp pairs are present in this nucleus. Figure from~\cite{Cruz-Torres:2019fum}.}
\end{center}
\end{figure}
\cite{Cruz-Torres:2019fum} computed the contact ratios, shown in Fig.~\ref{fig:nucl-contacts}, for a range of nuclei from $A=3$ to 40, using multiple $NN$ potentials including modern chiral EFT potentials (N2LO at different cutoff scales) and traditional phenomenological potentials (AV18, AV4'). They found that for spin-1 $np$ pairs, the ratio of the contact in nucleus $A$ to the contact in the deuteron is the same for all potentials, in both position and momentum space, for all nuclei studied up to $A=40$. Similarly, for spin-0 $pp$ pairs, the ratio of the contact in nucleus $A$ to the contact in $^4$He is potential-independent and consistent between position and momentum space calculations.

The GCF provides a compact way to describe SRC physics. It has been shown to describe the measured cross sections for $(e,e'p)$ and $(e,e'pp)$ events very well across a wide range of nuclei and kinematics, see~\cite{CLAS:2020mom}. It also connects nuclear SRC physics to the field of ultracold atomic gases, where an analogous Contact formalism was developed by \cite{Tan08a} and has been extensively tested.

\subsection{Nuclear Spectral Functions and Reaction Theory}\label{sec:theory:reaction}

The self-consistent Green's function (SCGF) method of \cite{Dickhoff:2004xx} computes the single-nucleon propagator $g(\omega)$ self-consistently: short-range correlations are included by summing particle-particle (ladder) diagrams to all orders in the self-energy, while long-range correlations are captured by particle-hole (ring) diagrams.  The resulting spectral function is fragmented over a broad range of energies, with $\sim 35\%$ of single-particle strength depleted from the mean-field peaks thereby making a robust prediction confirmed by $(e,e'p)$
measurements across many nuclei.  An important result is that this depletion is driven by both short-range correlations and long-range collective effects, and separating the two contributions requires detailed spectral-function calculations.

For two-nucleon knockout, $A(e,e'pN)B^*$, the cross section is sensitive to the two-body density matrix and provides the most direct probe of SRC pairs. Refs.~\cite{Vanhalst:2011es,Vanhalst:2012ur,Colle:2013nna} computed $(e,e'pp)$ and $(e,e'pn)$ cross sections using correlated nuclear spectral functions, demonstrating that the ratio of $pp$
to $pn$ emission rates directly reflects the $T=1$ to $T=0$ SRC ratio.  Because the tensor force strongly enhances $T=0$ ($pn$) correlations over $T=1$ ($pp$, $nn$) correlations, $pn$ pairs dominate two-nucleon knockout at high missing momenta. Final-state interactions between the ejected nucleons and the residual nucleus are treated via Glauber multiple-scattering
theory, and charge-exchange rescattering effects (which modify the isospin composition of
the detected pair) are accounted for explicitly. These reaction-theory calculations are essential for unambiguously extracting SRC pair properties from measured cross sections.

\section{SRCs and Connections to Other Topics}

\subsection{Neutron Stars}

Neutron stars are among the most extreme environments in the universe with about  1.4 solar masses compressed into a $\approx 10$-km sphere, gving central densities  several times larger than the nuclear saturation density $\rho_0\!\approx\!0.16\,{\rm fm^{-3}}$. At these densities, core  nucleons are at  separations  comparable to those of nucleons in SRC pairs.  Therefore the same physics responsible for SRC pairs in nuclei directly shapes the neutron star Equation of State (EOS), its chemical composition, and its cooling.

$np$ SRC dominance can dramatically affect the neutron star  proton momentum distribution. Inside a neutron star with a proton fraction of $\sim5-13\%$, $np$ pair dominance predicts that the probability to find a correlated proton with $k\ge k_F$ is significantly greater than in nuclei, see  \cite{frankfurt08}. 

These high-momentum protons  can change the standard neutron star cooling.  The direct URCA cooling processes $n\!\to\!p+e^-+\bar{\nu}_e$ and $e^-+p\!\to\!n+\nu_e$ rely on enough `holes' in the Fermi sphere to reduce the effects of Pauli blocking.  In the conventional Landau Fermi-liquid picture, after an initial period of direct URCA cooling, neutrino cooling of a cold neutron star is dominated by the modified URCA processes $n+N\!\to\!p+N+e^-+\bar{\nu}_e$ and $e^-+p+N\!\to\!n+N+\nu_e$. The much faster direct URCA processes  are forbidden by momentum conservation unless the proton fraction exceeds the threshold $Z/(A-Z)\!\geq\!1/8$ set by the triangle condition $k_F^p+k_F^e\!\geq\!k_F^n$, see~\cite{frankfurt08b}. This triangle condition assumes a Fermi gas momentum distribution. This implies that a high momentum proton in an SRC pair can satisfy the triangle condition by itself, opening a direct URCA channel even for $Z/(A-Z)\!<\!1/8$.  Above this proton-fraction threshold, the same high momentum nucleons enhance both the direct and modified rates of URCA cooling, see~\cite{frankfurt08b}.
The SRC-enhanced direct URCA cooling is   similar to the modified URCA cooling in that both rely on an $NN$ interaction  prior to beta decay. Thus, SRC pairing provides an experimentally tested mechanism for modified URCA cooling.

A complementary perspective is provided by the recent EOS analysis of~\cite{cai2025}, who track how the high momentum nucleons from SRCs modify the chemical composition of $\beta$-stable matter through the modified symmetry energy and how the Migdal--Luttinger $Z$-factor (describing the discontinuity of $n(k)$ at $k_F$) suppresses the neutrino emissivity of the direct URCA, modified URCA, $NN$ bremsstrahlung, and Cooper-pair breaking and formation processes (see~\cite{cai2025}. Two effects then act in opposing directions: the high momentum nucleons lower the critical proton fraction for direct URCA cooling (from the canonical $\sim\!11\%$ down to a few percent in some models, qualitatively consistent with~\cite{frankfurt08}), while the depletion $Z\!<\!1$ of the Fermi surface reduces each individual emissivity by powers of $Z_F^{n,p}$.  The net result is a strong, mass-dependent modification of the cooling curves, with massive stars preferentially activating the SRC-broadened direct URCA channel and cooling rapidly while lower-mass stars cool slowly through the $Z$-suppressed modified URCA, in good qualitative agreement with the observed spread of neutron-star surface temperatures, see~\cite{cai2025}.

The connection between short range correlation scattering measurements and neutron star phenomenology is presently qualitative. The same nuclear contacts and isospin structure of $n(k)$ that govern the cross sections discussed earlier in this review enter the nuclear EOS through the kinetic symmetry energy and the high momentum tail-modified pressure,
and they enter the neutrino luminosity through the proton occupation above $k_F$ and the Migdal--Luttinger $Z$-factor.  More work needs to be done to quantitatively connect SRC pairing in nuclei with neutron stars. 

\subsection{Neutrino Physics}
Short-range correlated $NN$ pairs can also be important for neutrino physics~\cite{cai2025}. By modifying the short-range two-body operators they can affect the matrix elements in neutrino-less double beta decay, see~\cite{Menendez2011,Menendez2014}, experimental measurements of the unitarity of the CKM matrix, see~\cite{Condren2022}, and neutrino-nucleus scattering, see~\cite{VanCuyck2016}.  

\section{Summary and Outlook}
Short-range correlated nucleon pairs are a fundamental feature of the nuclear ground state that lies beyond the independent-particle shell model. The key findings from the last few decades of experimental and theoretical work are:
\begin{enumerate}
    \item SRC pairs have been observed in both electron- and proton-nucleus scattering, showing that they are probe-independent.
    \item About 20\% of nucleons in medium to heavy nuclei belong to SRC pairs at any given instant. These nucleons carry almost all the high-momentum strength in the nuclear momentum distribution.
    \item SRC pairs are overwhelmingly $np$ pairs (approximately 90\%) at relative momenta of 300–-600~MeV/$c$. This ``$np$ dominance" is a direct consequence of the tensor part of the $NN$ interaction.
    \item At higher relative momenta above $\approx 600$~MeV/$c$, the $pp/np$ ratio increases toward the value expected from the central (scalar) force (i.e., simple pair counting). This transition from tensor to central correlations has been measured and is consistent with calculations across a wide range of $NN$ potentials.
    \item The center of mass momentum distribution of SRC pairs is approximately Gaussian-distributed in each Cartesian direction with a width of 100–-170~MeV/$c$, increasing with the nuclear mass. This is consistent with SRC pairs forming from  ordinary mean-field nucleons that temporarily fluctuate into a high-relative-momentum state.
    \item SRC pair abundances as measured by inclusive scattering ratios increase rapidly from deuterium to $^4$He, and then increase more slowly and then saturate from $^4$He to heavy nuclei. 
    \item The Generalized Contact Formalism provides an elegant theoretical description of SRC physics, allowing factorization of the momentum distribution into mean-field and SRC-pair parts. Contact ratios (relative SRC-pair probabilities) between nuclei are independent of the $NN$ potential used, showing that SRC probabilities are due to the long range part of the $NN$ interaction and SRC momentum distributions are due to the short-range part of the $NN$ interaction.
\end{enumerate}
The upgraded CEBAF accelerator at Jefferson Lab with electron beams up to 12~GeV has enabled many recent experiments to explore SRC physics. Future experiments at both Jefferson Lab and the planned Electron-Ion Collider (EIC) will enable new, higher-precision measurements of SRC pairs across the periodic table and at higher momenta. New radioactive ion beam facilities worldwide will allow SRC studies in unstable nuclei far from stability, connecting nuclear structure to astrophysical environments. Better theoretical tools including ab initio nuclear structure calculations and lattice QCD at the quark level will enable more rigorous interpretations of the data.

Short-range correlations remind us that even our best model of a complex system, the nuclear shell model, captures only part of the reality. The remaining 20\%, hidden in transient high-momentum fluctuations, reveals deep and universal properties of the strong nuclear force that continue to to be studied.

\bibliographystyle{Harvard}
\bibliography{references}

\end{document}